\documentclass[11pt]{article}
\pdfoutput=1
\usepackage[textwidth=15.2cm,textheight=22cm]{geometry}
\usepackage{amsmath,amssymb}
\usepackage{latexsym}
\usepackage{multicol}
\usepackage{datetime}
\usepackage{url}
\usepackage{simplewick}
\usepackage{color}
\usepackage{cite}

\usepackage{framed}
\usepackage{graphicx}
\usepackage{bm}
\allowdisplaybreaks[1]

\newcommand{\be}{\begin{equation}}
\newcommand{\ee}{\end{equation}}
\newcommand{\ba}{\begin{eqnarray}}
\newcommand{\ea}{\end{eqnarray}}
\newcommand{\bdm}{\begin{displaymath}}
\newcommand{\edm}{\end{displaymath}}

\newcommand\fr[1]{\frac{1}{#1}}

\newcommand{\dd}{\raisebox{10pt}{\tiny$\scriptscriptstyle\longleftrightarrow$}\hspace{-11.7pt}}
\newcommand{\db}{\raisebox{10pt}{\tiny$\scriptscriptstyle\longleftarrow$}\hspace{-11pt}}

\newcommand{\rom}[1]{\uppercase\expandafter{\romannumeral #1\relax}}

\def\b{\beta}
\def\a{\alpha}
\def\g{\gamma}
\def\s{\sigma}

\def\d{\delta}

\def\ba{\bar A}
\def\bb{\bar B}
\def\bc{\bar C}

\def\beq{\begin{equation}}
\def\eeq{\end{equation}}

\newcommand{\ie}{{\it i.e.\ }}

\newcommand{\ndt}{\noindent}

\newcommand{\D}{\mathcal{D}}

\def\bea{\begin{eqnarray}}
\def\eea{\end{eqnarray}}
\def\beas{\begin{eqnarray*}}
\def\eeas{\end{eqnarray*}}
\def\sla{\raise.15ex\hbox{$/$}\kern-.57em}

\def\d{\delta}

\def\spa#1.#2{\left\langle#1\,#2\right\rangle}
\def\spb#1.#2{\left[#1\,#2\right]}

\def\mD{\mathcal{D}}

\def\ul#1{{\underline{#1}}}

\def\da{{\dot{a}}}
\def\db{{\dot{b}}}
\def\dc{{\dot{c}}}
\def\dd{{\dot{d}}}

\def\bA{{\bar{A}}}
\def\bB{{\bar{B}}}
\def\bC{{\bar{C}}}

\def\ba{{\bar{a}}}
\def\bb{{\bar{b}}}
\def\bc{{\bar{c}}}
\def\bd{{\bar{d}}}
\def\be{{\bar{e}}}

\def\bda{{\bar{\da}}}
\def\bdb{{\bar{\db}}}

\def\bphi{{\bar{\phi}}}

\def\mO{\mathcal{O}}
\def\mN{\mathcal{N}}
\def\mD{\mathcal{D}}

\renewcommand{\th}{\theta}
\newcommand{\bth}{\bar\theta}

\def\a{\alpha}
\def\b{\beta}
\def\d{\delta}
\def\e{\epsilon}
\def\l{\lambda}

\def\s{\sigma}

\def\th{\theta}

\def\der{\partial}
\def\bea{\begin{eqnarray}}
\def\eea{\end{eqnarray}}

\begin{document}

\begin{titlepage}
\begin{flushright}
DIAS-STP-18-09
\end{flushright}
\vskip 1cm
\centerline{\LARGE{\bf{Towards a tensionless string field theory}}}
\vskip 0.3cm
\centerline{\LARGE{\bf{for the $\mN= (2,0)$ CFT in $d=6$}}}
\vskip 1.5cm
\centerline{Sudarshan Ananth$^\dagger$, Stefano Kovacs$^{\star}$, Yuki Sato$^{\sharp}$ 
and Hidehiko Shimada$^\diamond$}
\vskip 0.5cm
\centerline{${\,}^\dagger$\it {Indian Institute of Science Education 
and Research}}
\centerline{\it {Pune 411008, India}}
\vskip 0.3cm
\centerline{${\,}^\star${\it {Dublin Institute for Advanced Studies}}}
\centerline {\it {10 Burlington Road, Dublin 4, Ireland }}
\vskip 0.3cm
\centerline{${\,}^\sharp$\it {Department of Physics, Nagoya University}}
\centerline {\it {Chikusaku, Nagoya 464-8602, Japan}}
\centerline{\it{and}}
\centerline{\it {Department of Physics, Faculty of Science, 
Chulalongkorn University}}
\centerline {\it {Thanon Phayathai, Pathumwan, Bangkok 10330, Thailand}}
\vskip 0.3cm
\centerline{${\,}^\diamond$\it {Mathematical and Theoretical Physics Unit}}
\centerline{\it{Okinawa Institute of Science and Technology}}
\centerline {\it {1919-1 Tancha, Onna-son, Okinawa 904-0495 Japan}}
\vskip 1.5cm
\centerline{\bf {Abstract}}
\vskip .5cm
We describe progress in using the field theory of tensionless strings
to arrive at a Lagrangian for the six-dimensional
$\mathcal N=(2,0)$ conformal theory.
We construct the free part of the theory and propose an ansatz for the cubic vertex
in light-cone superspace.
By requiring closure of the $(2,0)$ supersymmetry algebra,
we fix the cubic vertex up to two parameters.
\vfill
\end{titlepage}

\numberwithin{equation}{section}

\section{Introduction}

The possible existence of a superconformal field theory with $(2,0)$ supersymmetry
in six dimensions was first pointed out in~\cite{RBNahm}. A 
string theory origin for such a conformal field theory (CFT) 
was proposed in~\cite{RBWitten} and the 
theory was then identified as a candidate for the description of the 
low-energy dynamics of M5-branes, important but elusive
degrees of freedom (DOF) in M-theory~\cite{RBStrominger}. In recent years, the 
theory has also played a crucial role in various developments in 
mathematical physics, with particular attention being devoted to the 
classification of BPS observables and the study of their properties both in six 
dimensions and, upon compactification, in lower 
dimensions~\cite{RBMooreLectures}.

The $\mathcal N=(2,0)$ theory is also interesting from 
the point of view of the theory space of quantum field theory. 
This space is governed by the renormalisation group 
flow~\cite{RBWilsonKogut} in which fixed points, 
\ie conformal field theories~\cite{RBPolyakovCFT}, are an 
essential feature. 
It is known that six dimensions is the highest dimension of 
spacetime that permits a theory with superconformal 
symmetries~\cite{RBNahm}. 
The very existence of a six-dimensional CFT is surprising 
because power-counting makes it difficult to write down interacting theories 
(except for a scalar $\phi^3$ coupling, which does not satisfy the requirement 
of positive definiteness of the energy) involving a dimensionless constant in 
dimensions higher than four.  

Despite the importance of the $\mathcal N=(2,0)$ theory and the attention it has attracted in
recent years, there is no consensus on whether it should admit a  
Lagrangian formulation. Various obstructions exist to the realisation of 
superconformal symmetry in a conventional six-dimensional local field theory. 
Several Lagrangian constructions have been proposed,
including the matrix model approach involving a low-energy 
limit~\cite{RBMatrixModelApproach1,RBMatrixModelApproach2}, the 
dimensional deconstruction
approach~\cite{RBDeconstruction},
and the decompactification limit of $d=5$ maximally supersymmetric 
Yang-Mills 
theory~\cite{RBDouglas, RBLambertPapageorgakisSchmidtSommerfeld}. 
For other proposals, see~\cite{RBLagrangianProposal1, RBLagrangianProposal2, 
RBLagrangianProposal3, RBLagrangianProposal4, RBLagrangianProposal5, RBLagrangianProposal6} 
and references therein. Another interesting approach is based on the idea of 
the conformal bootstrap~\cite{RBBootstrap}, which does not rely on the 
existence of a Lagrangian. 

Although the use of the bootstrap method may render a Lagrangian description
unnecessary, having an explicit Lagrangian formulation is 
desirable for a better understanding of the fundamental DOF
of the $(2,0)$ theory. Such a description would also
clarify the relationship of the $(2,0)$ CFT 
in $d=6$ to lower dimensional maximally supersymmetric theories and in 
particular the ${\cal N}=4$ super Yang--Mills (SYM) theory in four dimensions. 
Moreover, although the $(2,0)$ CFT is inherently non-perturbative, as implied 
by its M-theory origin, a Lagrangian description should make it possible to  
construct reliable weak-coupling approximation schemes valid in special 
sectors and/or for special observables, such as near-BPS quantities. 
These ideas were exploited  in~\cite{RBKSS1, RBKSS2} in the case of
the ABJM theory~\cite{RBABJM} -- the maximally supersymmetric
CFT in three dimensions, 
associated with coincident M2-branes -- which is also intrinsically strongly 
coupled. In~\cite{RBKSS1}, using the AdS/CFT correspondence, a 
perturbative analysis of the spectrum in a special sector of the ABJM theory 
was successfully compared to the dual AdS description provided by the 
pp-wave matrix model~\cite{RBBMN}.

In this paper we propose developing a Lagrangian for 
the $\mathcal N=(2,0)$ theory in six dimensions,
using String Field Theory (SFT) in 
light-cone gauge. The use of light-cone gauge is key to our approach since it 
allows us in principle to determine the interacting theory by a fairly 
straightforward -- albeit technically involved -- closure of the supersymmetry 
algebra~\cite{RBMandelstamSYM,RBBBBSuper}. 

It has been argued that the six-dimensional $(2,0)$ theory contains 
tensionless string DOF. In particular, in the M-theory 
construction in which the $(2,0)$ theory describes the low-energy dynamics 
of a collection of M5-branes, the strings arise from M2-branes stretched 
between M5-branes.
When the M5-branes are coincident the M2-branes reduce to 
closed strings in the world-volume of the M5-branes. Such strings are
tensionless as their tension is proportional to the (constant) M2-brane tension 
times the separation between the M5-branes. While of course this 
construction does not imply that the fundamental DOF
in the effective theory describing the world-volume dynamics of coincident 
M5-branes should be tensionless strings, it is certainly natural to consider 
such a possibility. 

In the case of the four-dimensional ${\cal N}=4$ SYM theory, open strings 
ending on $N$ coincident D3-branes give rise to matrix-valued point-like 
DOF. Similarly, when considering a stack of $N$ coincident
M5-branes, there are $N\times N$ configurations of M2-branes ending on
the M5-branes, with each cylindrical M2-brane degenerating to a closed string 
constrained to the six-dimensional world-volume of the M5-branes. Therefore 
we obtain a six-dimensional matrix-valued closed string theory, that we will 
formulate using the language of string field theory in light-cone gauge. 

The approach that we propose in this paper is to construct directly a theory 
of tensionless strings in six dimensions, using the light-cone string field theory 
formalism, rather than to take the tensionless limit in a theory with
tension. The main reason leading us to this choice 
is that the zero tension limit of an ordinary tensile string 
theory is problematic and not well understood~\footnote{
The zero tension limit of ordinary tensile string theory has been 
studied by many authors in connection with higher spin gauge theories.
For an overview and references see~\cite{RBSagnotti}.
The tensionless limit of bosonic covariant SFT~\cite{RBWittenSFT}
was studied in~\cite{RBBonelli}, where the possibility of formulating
the $(2,0)$ CFT as the zero tension limit of SFT was also mentioned.
Early work on tensionless strings includes 
\cite{RBSchild, RBLizziRaiSparanoSrivastava, RBKarlhedeLindstrom, RBLindstromSundborgTheodirisSuper, RBLindstromSundborgTheodirisSpinning, RBIsbergLindstromSundborg, RBIsbergLindstromSundborgTheodoridis, RBGustafssonLindstromSaltsidisSundborgVanUnge, RBAmorimBarcelosneto, RBGamboaRamirezRuizalbata1, RBGamboaRamirezRuizalbata2, RBBarcelosnetoRuizalbata}. 
Some discussions on the tensionless limit can be found in 
\cite{RBBagchiChakraborttyParekh} and references therein.
}.
This is analogous to the case of general quantum field theories, 
in which taking a zero mass limit often requires careful analysis. 
The most appropriate procedure to study such a limit would involve 
computing physical observables and then taking the limit on these. 
However, the conventional first quantised formulation 
of string theory, in our present understanding, only allows one to compute 
S-matrix elements, whereas the good 
observables in a conformal field theory such as the one we are trying to
construct are expected to be local correlation functions.
Since local correlators in tensile string theory are not understood
and, further, S-matrix elements in the tensionless limit 
can be singular and at least not straightforward to define,
we propose to construct the $(2,0)$ CFT directly as a 
tensionless string theory in six dimensions rather than trying to 
define it as the tensionless limit of some string theory with tension. 

The fact that the tensile strings and the $(2,0)$ CFT 
should have fundamentally different natural observables 
also supports our choice to use a second-quantised, string field theory, 
formulation~\footnote{The distinction between the first quantised and
the second quantised formulations is important at the interacting level.
For the free part, the two descriptions are directly
related to each other, in particular in the light-cone gauge.
}.  
This formalism should prove better suited to the study of the observables of a CFT. 
Further support for such 
an approach follows from the analogy with the case of point particles. 
The world-line (first quantised) formalism is not straightforward for the study 
of massless particles, which instead are simple to describe in the field theory 
(second quantised) language. 

Our approach may be compared to the standard treatment of Yang-Mills theory.
As is well known, it is easier to work with massless Yang-Mills theory directly, 
rather than thinking of it as a limit of a theory of massive interacting vector particles,
the essential reason being the gauge symmetry of the theory in the massless case.
One of course also uses the second-quantised field theory formalism, rather 
than a first-quantised formulation, for Yang-Mills theory.

A particular virtue of our approach regards the dimension of the coupling 
constant. In traditional field theory, the dimension of the coupling constant 
depends on the dimension of spacetime. 
This renders the program of writing down an interacting $d=6$ Lagrangian,
in particular with the correct supersymmetry, very difficult. 
In contrast, the physical dimensions of the coupling constant do not depend 
on the spacetime dimension in SFT and therefore, in principle, no obstruction 
arises from power counting arguments. We elaborate on this point in 
section~\ref{RSPowerCounting}.

Another promising feature in our proposal is related to dimensional reduction. 
The six-dimensional $(2,0)$ theory is expected to reduce to 
the $\mathcal N=4$ SYM theory in four dimensions when compactified on a 
torus. The coupling constant of the reduced theory, $g_{YM}$, is given by the 
formula $\frac{1}{g_{YM}^2}\sim \frac{R_1}{R_2}$, where $R_1$ and $R_2$
are the two compactification radii.
Although the dependence on $R_1$, in this formula, can be easily understood 
in terms of a standard Kaluza--Klein reduction, the dependence on $R_2$ 
is much harder to understand in the context of an ordinary 
local field theory. Using (tensionless) string DOF, on the other 
hand, means that wrapped strings play a role in the reduction, thus introducing
a distinction between the two compactification radii. This may lead to a 
mechanism for generating the required dependence on $R_2$ in the 
formula for the four-dimensional coupling constant. 

The choice of light-cone gauge allows one to focus exclusively on the 
physical DOF and in this gauge symmetry constraints can be 
more directly implemented, so that one can restrict or even determine the 
theory purely from symmetry considerations.
This idea of determining the interacting Hamiltonian by requiring the closure of
the symmetry algebra has proven extremely fruitful 
in the past~\cite{RBBBB, RBAkshayAnanth, RBAnanth, RBAnanthKarMajumdarShah}. 
In particular, the entire $\mathcal N=4$ SYM theory -- for which 
the light-cone superspace formulation was first obtained in~\cite{RBMandelstamSYM, RBBLN} -- 
can be  derived from closure of the superconformal algebra~\cite{RBABKR}.
The action describing light-cone superstring field theory in ten dimensions  
has also been derived to cubic order in this 
way in~\cite{RBGS1, RBGSB, RBGS2, RBGS3} and
the full Lorentz symmetry of the theory up to cubic order was verified 
in~\cite{RBLinden}.
We also recall that light-cone gauge bosonic string field theory was 
developed in~\cite{RBMandelstamCubicVertex, RBMandelstamLorentz, RBKakuKikkawa1, RBKakuKikkawa2, RBCremmerGervais1, RBCremmerGervais2, RBMarshallRamond} and 
a detailed study of the Lorentz invariance of the theory was presented in 
\cite{RBMandelstamLorentz, RBBengtssonLinden, RBKugo, RBSin, RBSaitohTanii1, 
RBSaitohTanii2, RBKikkawaSawada}. 

Our aim is to construct an interacting theory of tensionless strings having 
the right amount of supersymmetry and a dimensionless coupling constant 
(which is a necessary condition for the scale invariance of the model) in 
six space-time dimensions~\footnote{We expect that, as in the case of 
$\mN=4$ SYM in four-dimensions, the classical scale invariance
is not broken by quantum effects.}. 
In this paper, we present the construction of the 
quadratic and cubic parts of the SFT action. We formulate an ansatz, which 
we justify by using (part of) the restrictions imposed by the closure of the 
supersymmetry algebra. The cubic vertices that we obtain still contain two 
arbitrary parameters. 
Our construction is based upon the light-cone superspace 
formulation of the free particle with $(2,0)$ supersymmetry in six 
dimensions~\cite{RBRamond, RBAnanthBrinkRamonUnpublished}. 

Our approach combines features of both
the light-cone formulation of $\mN=4$ SYM and the supersymmetric closed 
SFT. It is similar to the former since our aim is to formulate a theory with 
tensionless (massless), matrix-valued DOF and sixteen
supercharges, while it resembles the latter because we are trying to construct 
a theory of closed strings.

This paper is organised as follows. In section \ref{RSSymmNotation},
we review the relevant symmetries of the theory and explain our notation,
with further details in appendices \ref{RSATensors} and \ref{RSASuperAlgebra}.
In section \ref{RSFree}, we introduce 
the string field, and give the free part, \ie the part 
which is quadratic in the string fields, of the symmetry charges.
In section \ref{RSOverlapInsertion}, we explain the notation
necessary for describing the cubic interaction part, and introduce the
two essential ingredients, the overlap and the insertion.
Section \ref{RSAnsatz} presents the ansatz for the cubic vertices, and
shows that the ansatz is consistent with the supersymmetry algebra.
A discussion of power counting is also presented.
In section \ref{RSConclusions} we conclude with a discussion.
Details involved in some of the definitions  and computations
are deferred to several appendices.

\section{Symmetries and notation}
\label{RSSymmNotation}

The theory we are interested in exhibits $\mN=(2,0)$ super-Poincar\'{e} symmetry
and its superconformal extension.
The associated R-symmetry is USp(4)~\cite{RBNahm, RBHoweSierraTownsend, RBStrathdee}.

We choose the metric with signature $(-, +, \ldots, +)$ and  introduce the 
light-cone coordinates
\begin{align}
x^+=\fr{\sqrt 2}\,(x^0+x^5)\, , \qquad x^-=\fr{\sqrt 2}\,(x^0-x^5)\ .
\end{align}
We denote the four transverse directions by $x^\a$, $\a=1,2,3,4$.
$x^+$ plays the role of time implying that $-P_+=P^-$ 
is the light-cone Hamiltonian. As is often done, we work on a surface 
defined by $x^+=0$. 

An SO(4) subgroup of the original SO(1,5) Lorentz symmetry, acting on the 
transverse directions $x^\alpha$, remains manifest. We introduce capital 
indices, $A,B,\ldots=1,2,3,4$, for the R-symmetry and lower case 
undotted and dotted indices, $a,b,\ldots$, ${\dot a},{\dot b}, \ldots=1,2$, to represent the 
SO(4)=SU(2)$\times$SU(2) spinor indices. 

The generators of the super-Poincar\'e algebra split into two 
varieties. The kinematical generators 
\begin{align}
P^+, Q_{KaA}, P_\alpha, M^{\a\b}, M^{+\a}, M^{+-}\ ,
\end{align}
which do not pick up corrections in the interacting theory, and the dynamical 
generators
\begin{align}
P^-, Q_{D{\dot a} A}, M^{-\alpha}\ ,
\end{align}
which do. When there is a possible ambiguity, such as in the case of the supercharges,
we use subscripts, $K$ and $D$, to differentiate between kinematical and dynamical 
generators. Dynamical generators transform fields non-linearly, while 
kinematical generators act linearly on the fields. In this light-cone formalism, the 
super-Poincar\'e algebra imposes strong constraints on the theory, including on 
the Hamiltonian, $P^-$. These symmetry algebra constraints are what we will 
use to determine the interacting Hamiltonian.
The entire super-Poincar\'e symmetry algebra is presented in 
appendix~\ref{RSASuperAlgebra}.

We will not consider the closure of the full superconformal algebra
and will instead focus on just the super-Poincar\'e part of the algebra. 
We believe that this part of the superalgebra, together with the requirement 
of a dimensionless coupling constant, is sufficient to determine the ansatz.
It would also be interesting to examine the entire superconformal symmetry, 
as was done previously for $\mN=4$ SYM~\cite{RBABKR}.

\section{The free theory}
\label{RSFree}

Our study of the free theory begins with the superfield functional
\begin{equation}
\phi^I_{P^+} [x^\a(\s), \th^{aA}(\s)]\ .
\label{superfield}
\end{equation}
We do not write the dependence on the time coordinate $x^+$ explicitly.
The string field depends on the total momentum $P^+$ and not on 
the momentum density $p^+(\s)$, because 
the choice of the light-cone gauge condition implies that $p^+(\s)$ does not  
depend on $\s$~\cite{RBGGRT}.
The fermionic coordinates $\th^{aA}$ carry both R-symmetry and
SO(4) spinor indices.
As explained in the introduction, 
we expect to have $N\times N$ matrix-valued string fields when we have $N$ M5-branes.
We use indices $I, J, \ldots$ to label these matrix DOF. 
We will later fix a Lie algebra and assume $I, J, \ldots$ to be 
Lie algebra indices running from $1$ to the dimension of the Lie algebra.
The $\s$ coordinate takes values in an interval of length $[\s]$. We choose
\begin{align}
-[\s]/2 < \s < [\s]/2\ .
\end{align}
The length $[\s]$ is taken to be proportional to $P^+$ 
and the coefficient of proportionality is denoted by $p^+$, \ie
\begin{align}
\frac{P^+}{[\s]} = p^+\ .
\end{align}
$p^+$ is a conventional constant and it is a c-number (it commutes with everything).
The fermionic coordinates $\th^{1A}$ and $\th^{2A}$ are related by 
complex conjugation,
\begin{equation}
\overline{\th^{aA}}
=B^{\ba}{}_b B^{\bar{A}}{}_B
\th^{b B},
\end{equation}
where $B^{\ba}{}_b$ is proportional to the $\e$-tensor.
For our definition of tensor structures such as the $B$'s
associated with the light-cone little group SO(4)
and the R-symmetry group USp(4), see appendix \ref{RSATensors}.
We will refer to $\th^{1A}$ as $\th$ and $\th^{2A}$ as $\bth$ below when appropriate.

\subsection{Chiral derivatives, supersymmetry and level-matching}
There are two different formulations of supersymmetric theories
in terms of light-cone superfields. 
In one approach, one uses superfields which depend only on $\th$ (or $\bth$).
For $\mN=4$ SYM in four dimensions, this approach was introduced in~\cite{RBMandelstamSYM}.
The formulation of spacetime supersymmetric SFT by 
Green, Schwarz and Brink~\cite{RBGS1, RBGSB, RBGS2, RBGS3} also belongs to this 
class of models.
In the other approach, one uses superfields depending on both $\th$ and $\bth$,
and certain chirality constraints are imposed, 
as was done for $\mN=4$ SYM in~\cite{RBBLN}.
While the former choice has the advantage of being direct, 
in the latter, formulae for the charges and the
power-counting procedure~\cite{RBBLNFinitess} 
are more transparent because  
fermionic coordinates enter in supercovariant combinations.

We adopt the latter approach. Our superfields depend on both $\th$ and $\bth$,
\ie $\th^{1A}$ and $\th^{2A}$. We  
impose the fundamental chirality constraint
on our superfield for each value of $\s$, 
\begin{equation}
d_{1 A}(\s) \phi= 0\ , 
\label{eq:chiralityderivative}
\end{equation}
where the chiral derivative is defined by
\begin{align}
d_{ a A}(\s)=\frac{\d}{\d \th^{a A}(\s) }+\frac{p^+}{\sqrt{2}} \th^{bB}(\s) \e_{b a} C_{BA}\ .
\end{align}
$C_{BA}$ is defined in appendix \ref{RSATensors}.

One can solve the constraint (\ref{eq:chiralityderivative}),
\begin{equation}
\phi_{P^+}(x^\a, \th, \bar{\th})=
e^{\frac{1}{\sqrt{2}} p^+ \int \th^A \bar{\th}_A d\s }
\Psi_{P^+}(x^\a, \bth)\ .
\end{equation}
Here
$\Psi$ 
is an arbitrary superfield depending only on $\bth$, which
can be identified with the superfield in an approach
analogous to \cite{RBMandelstamSYM, RBGS1, RBGSB, RBGS2, RBGS3}.

The superstring field is a natural extension of
the superfield for a superparticle in six-dimensional spacetime constructed in 
\cite{RBRamond, RBAnanthBrinkRamonUnpublished}.
If one focusses on the dependence of the string field on the
zero-mode part of $x(\s)$ and $\th(\s)$, one obtains the superfield for the superparticle
(for each value of the index $I$).
The superfield 
corresponds to the tensor multiplet~\cite{RBHoweSierraTownsend} of $(2,0)$ supersymmetry,
and gives the light-cone superfield corresponding to the $\mN=4$ SYM 
theory in four-dimensions~\cite{RBBLN} upon dimensional reduction.
This gives additional support to our idea that the superstring field is a natural choice 
for the construction of the $(2,0)$ theory~\footnote{
In the degenerate case of a single M5-brane~\cite{RBClausKalloshVanProeyen}, the 
$(2,0)$ CFT is conventionally believed to be a free theory of fields
belonging to the tensor multiplet. 
Putting $N=1$ in our case also leads to a free theory
with very many light degrees of freedom including the
tensor multiplet associated with the zero mode.
There is no immediate contradiction here since,
being free, these fields are completely decoupled.
}.
In particular,  it incorporates the self-duality property of the theory,
because the tensor multiplet includes a two-form gauge field with 
self-dual field strength.
Although our formulation is based on closed string DOF, 
it is nevertheless non-gravitational since 
the tensor multiplet does not contain any field of spin 2.

We introduce the local supersymmetry generator
\begin{align}
q_{a A}(\s)
=
\frac{\d}{\d \th^{a A}(\s) }
-
\frac{p^+}{\sqrt{2}} \th^{bB}(\s) \e_{b a} C_{BA}\ ,
\label{RFDefq}
\end{align}
which satisfies the following anti-commutation relations
\begin{align}
[
q_{a A}(\s), q_{bB}(\s')
]
=&-\sqrt2 p^+ \e_{ab} C_{AB} \d(\s-\s')\ ,
\\
[q_{aA}(\s), d_{bB}(\s')]=&0\ ,
\\
[
d_{a A}(\s), d_{bB}(\s')
]
=&\sqrt2 p^+ \e_{ab} C_{AB} \d(\s-\s')\ .
\end{align}
Here and in the rest of the paper we use square brackets to denote both
commutators and anti-commutators, depending on the Grassmann parity of the operators
involved.
We also define
\begin{align}
p_\a(\s) = -i \frac{\d}{\d x^\a(\s)}\ .
\end{align}

A level matching condition should be imposed on the string fields, which ensures
that the state be invariant under shifts of $\s$. The condition is related to the
requirement of global existence of $x^-$,
\begin{align}
\int \frac{\der x^{-}}{\der \s} d\s = 0\ ,
\end{align}
where the bosonic contribution to $\der_\s x^-$ is \cite{RBGGRT}
\begin{align}
\frac{\der x^-}{\der \s} =\frac{1}{p^+}p_\a \frac{\der x^\a}{\der \s}\ . 
\end{align}
When fermionic DOF are incorporated, 
the level matching condition becomes
\begin{align}
\left(
\int
\left(p_\a \frac{\der x^\a}{\der \s} 
-
i  \frac{\der \th^{aA}}{\der \s} (\s) \frac{\d}{\d \th^{aA}(\s)} 
\right) d\s
\right)
\phi =0
\label{RFLevelMatching}
\end{align}
and we have
\begin{align}
\frac{\der x^-}{\der \s}
=
\frac{1}{p^+}
\left(  p_\a \frac{\der x^\a}{\der \s}
-
i  \frac{\der \th^{aA}}{\der \s} (\s) \frac{\d}{\d \th^{aA}(\s)} 
\right)\ , 
\label{RFDefXMinus}
\end{align}
which defines $x^-(\s)$ up to the zero-mode part
\begin{align}
X^-=\frac{1}{[\s]} \int x^-(\s) d\s\ .
\end{align}

\subsection{Generators}

We are now in a position to write down 
the ``free" part of the various generators in our algebra. 
To simplify our presentation, we will use the language of the 
first quantised theory: we present the various charges as
operators acting on the string fields. 
The charges in the second quantisation formulation can be written down
basically by sandwiching the first quantised charge between $\bphi$ and $\phi$
in the usual way.

We begin by noting that the fist-quantised Hamiltonian for the
tensionless string in the light-cone gauge is simply
\begin{align}
P^-=\int \fr{2p^+}\,{(p^\alpha(\s))}^2\,d\s\ ,
\label{RFPMinus}
\end{align}
and does not contain the usual $(\der_\s x^\a)^2$ term
which is proportional to the square of the tension~\cite{RBGGRT}. 
This formula is unchanged even if one includes fermionic DOF.
Equation (\ref{RFPMinus}) shows that, while an ordinary tensile string can be understood as 
a collection of harmonic oscillators, 
a tensionless string is a collection of free particles. 
Each part of the string moves independently and all terms involving
$\der_\s$ vanish, except for the important level matching conditions (\ref{RFLevelMatching})
and the associated formula for the $x^-$ coordinate (\ref{RFDefXMinus}).
This makes the construction of the generators (except for $M^{-\a}$) quite easy;
we can start from the superparticle case~\cite{RBRamond, RBAnanthBrinkRamonUnpublished} 
and we can then simply add the $\s$-dependence.
These properties may be considered as a direct realisation 
of the idea of string bits~\cite{RBStringBits}.

For the supersymmetry charges we have
\begin{align}
Q_{KaA}&=\int q_{aA}(\s) d\s\ ,
\end{align}
\begin{equation}
Q_{D\da A}=\int
\frac{1}{\sqrt2}
q_{ b A}(\s) \frac{1}{p^+} \e^{bc} p^\a(\s) \s^\a{}_{c\da}
d\s\ . \label{RFQDZero}
\end{equation}
Other Poincar\'e generators include
\begin{equation}
M^{+\a} = \int - x^\a(\s) p^+ d\s
=
- X^\a P^+\ ,
\label{RFMPlusAlpha}
\end{equation}
\begin{equation}
M^{\a\b}
=\int \Big[
x^\a(\s) p^\b(\s) - x^\b(\s) p^\a(\s)
-
i \frac{\sqrt2}{8} \frac{1}{p^+}
\s^{\a\b a}{}_c \e^{cb}
C^{-1 A B} q_{a A}(\s) q_{b B}(\s) \Big]
d\s\ ,
\label{RFMAlphaBetaFirstQ}
\end{equation}
and
\begin{equation}
M^{+-} = - \frac12\left(X^-P^+ + P^+ X^- \right) - \int \frac{i}{2} \th^{a A}(\s) \frac{\d}{\d \th^{a A}(\s)} d\s\ .
\label{RFMPlusMinusFirstQ}
\end{equation}
All three Lorentz generators in (\ref{RFMPlusAlpha})-(\ref{RFMPlusMinusFirstQ})
are kinematical. The only dynamical Lorentz generator is
\begin{align}
\begin{split}
M^{-\a}=&\int \Big[
x^-(\s) p^\a(\s) - \frac12 \left(x^\a(\s) p^-(\s)+ p^-(\s) x^\a(\s)\right)
+\frac{i}{2}  \th^{aA}(\s) \frac{\d}{\d \th^{aA}(\s)} \frac{p^\a(\s)}{p^+} \\
&+
\frac{\sqrt{2}}{8} i
\frac{p^\g(\s)}{(p^+)^2}
q_{aA}(\s)
q_{bB}(\s)
\s^{\a\g}{}^{ab}
C^{-1 AB} \Big]
d\s\ .
\end{split}
\label{RFMMinusAlphaFirstQ}
\end{align}
The algebra satisfied by these generators is presented in appendix \ref{RSASuperAlgebra}. 
We have explicitly verified the commutators 
without taking care of ordering issues in the definition of 
products of operators, \ie
only at the level of the Poisson brackets.
Useful formulae and an outline of the computation of 
the commutator $[M^{-\a}, M^{-\b}]$ 
are presented in appendix \ref{RSAMM}.

The action of the charges 
on the superfield does not spoil 
the chirality constraint (\ref{eq:chiralityderivative}) because 
the charges are written in terms of $q$'s which anti-commute with 
chiral derivatives, $[q(\s), d(\s')]=0$.
For $M^{+-}$ and $M^{-\alpha}$, which contain $\th$ and $\frac{\d}{\d\th}$ explicitly, 
the consistency with the chirality constraint needs to be checked.
Using arguments similar to those in appendix \ref{RSAMM}, 
one can show 
\begin{align}
[M^{+-},d_{aA}(\sigma)] 
&=  \frac{i}{2} d_{aA} (\sigma) - i \partial_{\s}( \s  d_{aA} (\s))\ , 
\label{RFMpmD}\\
[M^{-\a}, d_{aA}(\s)] 
&= - \frac{i}{2}\frac{p^\a (\s)}{p^+} d_{aA}(\s) 
+ i \partial_{\s} \left( 
\left( 
\int^{\s}_{-[\s]/2} p^{\a}(\s')\ d\s' 
- \frac{P^{\a}}{2} 
\right) 
d_{aA}(\s) 
 \right)\ , 
 \label{RFMmaD}
\end{align}
as a consequence of the fact that $d_{aA}$ transforms as a density.
This yields 
\begin{equation}
[M^{+-},d_{aA}(\sigma)] \phi = 0\ , \ \ \ 
[M^{-\a}, d_{aA} (\s)] \phi =0\ ,
\end{equation}
which assures the consistency of the action of the generators with the chirality constraint.  

\section{The interacting theory: overlap and insertions} 
\label{RSOverlapInsertion}

We now wish to introduce interactions in this formalism with the focus being 
on cubic interactions. We label the three strings using indices 
$r,s,\ldots=1,2,3$. 
String $3$ is chosen to be the long one with strings $1$ and $2$ connecting 
to it or string $3$ splitting into $1$ and $2$. The range of $\s_1, \s_2, \s_3$ is 
denoted by $[\s_1], [\s_2], [\s_3]$ respectively. We require that
\begin{equation}
[\s_1]+[\s_2]=[\s_3]\ ,
\label{RFSquareBracketSigmaConserv}
\end{equation}
which also follows from the fact that $[\s]$ is proportional to
the conserved momentum $P^+$, so that (\ref{RFSquareBracketSigmaConserv}) is equivalent to
\begin{equation}
P^+_1+P^+_2=P^+_3\ .
\label{eq:momentumconservation}
\end{equation}
It is convenient to introduce $\s$ which takes value in the whole interval 
$I=I_3$. The whole interval $I$ consists of two ``intervals'' $I_1$ and $I_2$ 
respectively for strings $1$ and $2$. We use the following scheme
\begin{equation}
I=I_3=[-[\s_3]/2,[\s_3]/2]\ ,
\end{equation}
\begin{equation}
I_1=[-[\s_1]/2,[\s_1]/2]\ ,
\end{equation}
\begin{equation}
I_2=[[\s_1]/2, [\s_3]/2]+
[-[\s_3]/2, -[\s_1]/2]\ .
\end{equation}
Each $\s_r$ takes values within
$[-[\s_r]/2,[\s_r]/2]$
for $r=1,2,3$. 
$\s$ and $\s_r$ $(r=1,2,3)$ are related by
\begin{equation}
\s_3=\s\ ,
\end{equation}
\begin{equation}
\s_1=\s \mbox{ for $\s \in I_1$}\ ,
\end{equation}
\begin{equation}
\s_2=\s-[\s_3]/2 
\mbox{ or }
\s_2=\s+[\s_3]/2 
\mbox{ for $\s \in I_2$}\ .
\end{equation}
\begin{figure}[h]
\centering
\includegraphics[width=0.65\textwidth]{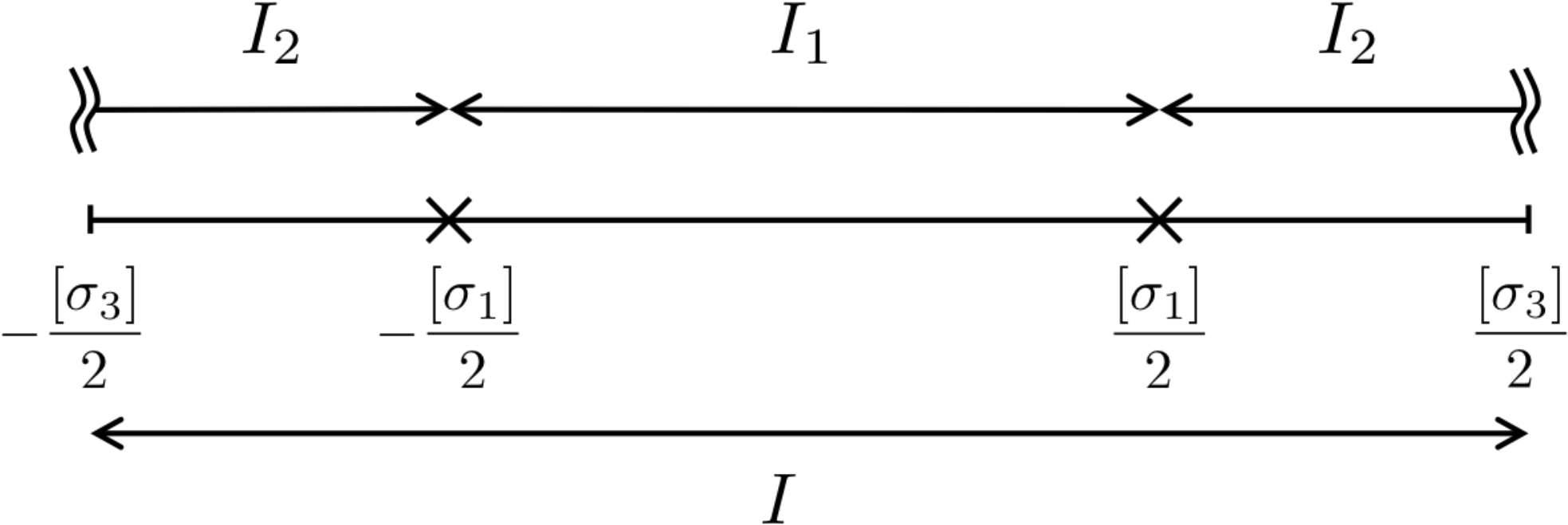}
\caption{
The $\s$-coordinates of closed strings $1,2$ and $3$ are 
defined on intervals $I_1, I_2$ and $I=I_3$.
The crosses indicate the interaction point.
}
\label{interval}
\end{figure}

Following the work on superstring theory in the spacetime supersymmetric
formalism \cite{RBGS1, RBGSB, RBGS2, RBGS3}, 
we introduce the two building blocks used to construct 
the cubic interactions: the overlap and the insertions. 
The overlap 
is a delta functional connecting the third string to the first and second strings. 
Local insertions of operators at the interaction point are also necessary.
These same ingredients (the overlap and the insertions) 
can be defined in the tensionless case as well. 

As usual, it is easier to work with discretely labelled variables
by introducing mode expansions. We introduce the Fourier components of $x_r(\s_r)$ by
\begin{equation}
x_r(\s_r)=\sum_n x^{r n} 
e^{i n \frac{2\pi}{[\s_r]} \s_r}\ . 
\label{RFFourierOfX}
\end{equation}
The canonical conjugate of $x^n$,  $p_n$, is 
\begin{equation}
p_{r n}=
\int p_r(\s_r) 
e^{i n \frac{2\pi}{[\s_r]}\s_r}
d\s_r 
\end{equation}
and $p_{r0}$ is the total transverse momentum $P_{r}$ (we omit $\a$ indices). 
The Fourier modes for $r=1,2$ and for $r=3$ respectively define two sets of  basis vectors.
We define a matrix $A$ relating the basis associated with the third string
to that associated with the first and second strings by
\begin{equation}
x^{rn} = A^{rn}{}_{3m} x^{3m} \quad (r=1,2)\ .
\label{RFDefA}
\end{equation}
We have
\begin{equation}
A^{rn}{}_{3m}
=
\frac{1}{[\s_r]}
\int_{\s\in I_r}
e^{-i \frac{2\pi}{[\s_r]}n \s_r}
e^{+i \frac{2\pi}{[\s_3]}m \s_3}
d\s\ .
\end{equation}
The overlap for the bosonic DOF is expressed as 
\begin{equation}
V_B= \prod_{r=1,2} \prod_n \d(x^{rn} - A^{rn}{}_{3m} x^{3m})\ .
\label{eq:v}
\end{equation}
For the fermionic component, we use
\begin{align}
V_F=
\prod_{r=1,2}\prod_{a=1,2}\prod_n \delta (\theta^{ran}-A^{rn}{}_{3m}\theta^{3am})\ .
\end{align}

Our philosophy in this paper is very similar, in spirit, to that followed in~\cite{RBABKR}. 
In order to build a consistent interacting theory,
we start with an ansatz for the dynamical supersymmetry generators. 
We allow the entire symmetry algebra to constrain our ansatz 
and finally use the fact that 
the Hamiltonian for the interacting theory can be written as 
the ``square" of the dynamical supercharge.

In general, the delta function (the overlap) is not
sufficient to construct the dynamical charges in light-cone gauge field theory and
one has to ``insert'' operators such as derivatives in $x$ 
and their fermionic counterparts acting on the overlap part.
This is the case both for $\mN=4$ SYM in four dimensions~\cite{RBMandelstamSYM, RBBLN} 
and for superstring field theory~\cite{RBGS1, RBGSB, RBGS2, RBGS3}.
In string theory it is not possible to insert the operators at an arbitrary point in $\s$.
The insertion should only act at the interaction point.

The insertion operator we choose is represented by 
the functions $w_r(\s)$ $(r=1,2)$, which have delta function
like singularities at the interaction point,
\begin{align}
w_1(\s_1)=&\d\left(\s_1-\frac{[\s_1]}{2}\right)
=\d\left(\s_1+\frac{[\s_1]}{2}\right)\ , 
\label{RFDefwFirst}
\\
w_2(\s_2)=&-\d\left(\s_2-\frac{[\s_2]}{2}\right)
=-\d\left(\s_2+\frac{[\s_2]}{2}\right)\ ,
\label{RFDefwSecond}
\end{align}
where we assume that the delta functions satisfy appropriate periodicity conditions.
In the mode number representation, we have
\begin{align}
w^{1n}=&\frac1{[\s_1]} (-1)^n\ ,
\\
w^{2n}=&\frac1{[\s_2]} (-1)^{n+1}\ .
\label{RFDefwLast}
\end{align}
The rationale for this choice is described in 
appendix~\ref{RSAInsertion}.

Now that we have an overlap and an insertion, we are in a position 
to write down an ansatz for the dynamical supersymmetry generator, 
describing a cubic interaction between the tensionless string fields. This is the 
focus of the next section.

\section{Ansatz for cubic interaction terms}
\label{RSAnsatz}

In general dynamical charges have an expansion, which in the case of 
$Q_{D}$, for instance, takes the form
\begin{align}
Q_{D}^{(0)}+Q_D^{(1)}+Q_D^{(2)}+\cdots \, .
\end{align}
Here $Q_D^{(0)}$ is the free part, quadratic in the string fields,
$Q_D^{(1)}$ is the cubic interaction part, containing three string fields, and so 
forth. The form of the ansatz is chosen so as to satisfy the 
super-Poincar\'{e} algebra (listed in appendix \ref{RSASuperAlgebra})
order by order in terms of the number of fields
involved. 
The cubic part of a dynamical charge consists of two terms 
respectively involving $\overline{\phi} \overline{\phi} \phi$ and $\overline{\phi} \phi \phi$.
Since one of them can be easily recovered 
from the other by the hermiticity conditions presented in appendix \ref{RSASuperAlgebra},  
we will hereafter only write the $\overline{\phi} \phi \phi$ part.
Our ansatz for $Q^{(1)}_D$ is
\begin{align}
Q^{(1)}_{D\dot a A}
&= f^I{}_{JK} 
\int 
\overline{\phi_{P^+_3}}_I [x_3,\theta_3] \notag \\
& \ \ \ \times  
\left( 
(p_{\alpha} \cdot w) (\sigma^{\alpha b}{}_{\dot a} d_{bA} \cdot w) 
(p^+)^{\lambda_0}
(P^+_1)^{\lambda_1}
(P^+_2)^{\lambda_2}
(P^+_3)^{\lambda_3} 
\delta (P^+_1 + P^+_2 - P^+_3) V     
\right) \notag \\ 
& \ \ \ \times 
\phi_{P^+_1}{}^J[x_1,\theta_1] \phi_{P^+_2}{}^K[x_2,\theta_2]
\prod^3_{r=1} 
dP^+_r \mathcal{D}\theta_r  \mathcal{D}x_r  \ .
\label{RFQDOne}
\end{align}   
Here we assume that 
$f^{I}{}_{JK}$ are the structure constants of a Lie algebra.
For the case of $N$ M5-branes in flat spacetime, $f^{I}{}_{JK}$
should correspond to U($N$).
$\lambda_0, \cdots , \lambda_3$ are parameters to be determined.
Below we will partially fix them by requiring 
invariance under rescaling of the $\s$ coordinate
and using power counting arguments. In (\ref{RFQDOne})
$V= V_BV_F$ and   
\begin{align}
&p_{\alpha} \cdot w 
= \int  p_{\alpha} (\sigma) w(\sigma) d \sigma
= p_{\alpha rn} w^{rn}\ , \\
&d_{bA} \cdot w = \int d_{bA}(\sigma) w(\sigma) d\sigma\ 
= d_{bA rn} w^{rn}\ .   
\end{align} 
The form of the ansatz is fixed basically by requiring that it
has the correct index structure.

If one exchanges $r=1$ and $r=2$ and the dummy indices $J, K$ in the above formula,
the result will have $\l_1$ and $\l_2$ exchanged. Furthermore one has a factor of $-1$
from each $w$ (compare (\ref{RFDefwFirst})-(\ref{RFDefwLast})) and a factor of $-1$ from $f$.
This implies that we must avoid choosing 
$\l_1=\l_2$ to have a non-vanishing ansatz. 
The ansatz for $P^{- (1)}$ will be determined below from the supersymmetry algebra.

\subsection{Power counting in SFT}
\label{RSPowerCounting}

We briefly discuss the power counting analysis of the cubic vertex.
The first step is to notice that the 
appearance of $\th$ and $\bth$ is accompanied
by factors of $p^+$, so that the integral measure for the fermionic
coordinates is dimensionless~\footnote{
This is because of the 
anti-commutation relations in superstring theory in the
light-cone gauge, $[\th(\s), \bth(\s')]\sim \frac{1}{p^+}\d(\s-\s')$.}.
The fermionic coordinates only contribute to the physical dimensions
through $q$'s and $d$'s and we will omit the dependence on the $\th$ coordinates 
of the string field in this subsection.

The dimension of the string fields turns out to be infinite.
We thus introduce a regularisation
where we discretise the $\s$ variables by $M$ 
string bits 
\begin{equation}
\phi_{P^+}(x_1^\a, \cdots, x_M^\a)\ .
\end{equation}
The dimension of the string field can be determined
by noting that it can be considered as the wave function of the 
first-quantised string theory~\footnote{
We note that in general one can redefine the string field
by multiplying it by factors of $P^+$. We do not introduce such a redefinition.
This choice is related to shifts of the operators
$M^{+-}$ and $M^{-\a}$ in (\ref{RFMPlusMinusFirstQ}) and (\ref{RFMMinusAlphaFirstQ})
and it is reflected in the hermitian ordering between
$X^-$ and $P^+$ and $p^-$ and $x^\a$ respectively.}.
Thus the normalisation factor
\begin{equation}
\int |\phi_{P^+}(x_1^\a, \cdots, x_M^\a)|^2 dP^+ d^4x_1 \cdots d^4x_M\ ,
\end{equation}
should be dimensionless implying that  
the string field $\phi$ has dimension
\begin{equation}
[\phi]=\frac{1}{2}\times (4M-1)\ ,
\end{equation}
which depends on the number of bits.

In the bit representation, the overlap delta functional $V$ is 
\begin{align}
V=
\prod_{n=1}^{M_1} \d\left(x_{3 n}-x_{1 n}\right)
\prod_{n'=1}^{M_2} \d\left(x_{3 (M_1+n')}-x_{2 n'}\right)\ .
\end{align}
The schematic form
(omitting factors irrelevant to the power counting) of 
the supercharge $Q_D^{(1)}$,
after carrying out the $\mD x_1 \mD x_2$ integrals using the delta functions, 
is
\begin{align}
Q_D^{(1)}\sim & \int \prod^{M_3}_{n=1} d^4 x_{3n} 
dP_1^+
dP_2^+
dP_3^+
\overline{\phi_3} 
\phi_1
\phi_2
\nonumber
\\
&
p\cdot w q\cdot w 
\d( P_1^+ + P_2^+ -P_3^+)
\left(p^+\right)^{\lambda_0}
\left(P_1^+\right)^{\lambda_1}
\left(P_2^+\right)^{\lambda_2}
\left(P_3^+\right)^{\lambda_3}\ . 
\label{RFCubicVertexPowerCounting}
\end{align}
We note that we are not introducing any dimensionful coupling constant here;
this is a requirement we impose on the SFT in order to construct a scale invariant theory.

Requiring that both sides of (\ref{RFCubicVertexPowerCounting}) have dimension $\frac12$,
we find 
\begin{equation}
\l_0+\l_1+\l_2+\l_3=-\frac{3}{2}\ .
\label{RFConstrLambdaFromDimAnalysis}
\end{equation}

An essential feature in the power-counting analysis of SFT presented above 
is that the $M$-dependent term in the dimension arising from the string fields
\begin{align}
[\phi_1]+[\phi_2]+[\phi_3]
=
\frac{1}{2}\times (4M_1-1)
+
\frac{1}{2}\times (4M_2-1)
+
\frac{1}{2}\times (4M_3-1)\ ,
\label{RFPowerCountingStringFields}
\end{align}
is exactly cancelled by the dimension of the measure
\begin{align}
[\prod_{n=1}^{M_3} d^4 x_{3 n}] = -4M_3\ ,
\label{RFPowerCountingMeasure}
\end{align}
because of the conservation of the number of bits 
\begin{equation}
M_1+M_2=M_3\ ,
\label{RFBitConservation}
\end{equation}
for the cubic vertex.

We observe that this cancellation implies that 
the dimensional analysis is independent of the number of transverse directions,
as can be seen from (\ref{RFPowerCountingStringFields}) and (\ref{RFPowerCountingMeasure}).
This is in sharp contrast with the dimensional counting in 
traditional field theories.
The power counting in SFT is favourable 
compared to that in usual QFT in this sense.

In the SFT case under consideration the free part of the action 
contains a term which schematically can be written as
\begin{align}
\int 
\overline{\phi_{P^+}}(x_1, \ldots, x_M)
\sum_{n=1}^M
\left(\frac{\der}{\der x_n^\a}\right)^2
\phi_{P^+}(x_1, \ldots, x_M) 
dP^+ d^4 x_1 \cdots d^4 x_M\ .
\end{align}
Comparing this formula in the case $M=M_3$ 
to the cubic vertex (\ref{RFCubicVertexPowerCounting}) we see that
the terms quadratic and cubic in the fields essentially have the same structure;
the difference only lies in how we group the string bits into different string fields.
This is the origin of why the power counting analysis does not depend
on the spacetime dimension.
This in turn reflects the basic feature of string theory that
locally string interaction and string propagation cannot be distinguished.

This result may have been expected as it is well known  
that the coupling constant in string theory is dimensionless
irrespective of the spacetime dimension. 
The property of possessing a dimensionless coupling constant 
potentially makes tensionless string theory
a natural framework for constructing theories with conformal symmetry.

The parameter $\l_0$ is fixed considering 
a rescaling of the $\s$ coordinate under which $[\s]$ becomes $\a [\s]$. 
Under this transformation $p_\a$ turns into $p_\a/\a$, \ie it transforms as a density. 
$p^+$, $d_{bA}(\s)$, and $w(\s)$ are also densities.
Taking into account the two $\s$ integrals involved in 
the definition of $p\cdot w$ and $q\cdot w$, we see that
\begin{align}
\l_0=-2\ .
\end{align}
Combining this with (\ref{RFConstrLambdaFromDimAnalysis}), we have
\begin{align}
\l_1
+
\l_2
+
\l_3
= \frac{1}{2}\ .
\label{RFConstrLambdaFromDimAnalysisFinal}
\end{align}

\subsection{Computation of commutators} 
\label{RSCommutatorComputation}
We explicitly work out the commutators 
to show that the ansatz is consistent with the superalgebra.

An issue in the computation is the potential singularity which
can occur because of the multiplication of 
operators at the same point in $\sigma$-space. 
To perform the computations in a well defined manner we use
a regularisation scheme,
analogous to that introduced in \cite{RBMandelstamLorentz},
in which operators are smeared.
For most of the commutators a computation done using smeared operators,
in the limit $\e\rightarrow 0$ (where $\e$ is the length scale 
associated with smearing),
produces a result which is identical to 
that of a formal computation without regularisation.
For the computation of the commutator $[Q_D, P^{-}]$, however, 
smearing makes a difference. Also, it is necessary to evaluate 
the result of the computation, which includes delta functionals,
by means of test functionals.
In this section, we avoid the explicit introduction of smearing.
Details regarding smearing and test functionals 
are discussed in appendix \ref{RSASmearingTestFunctional}.

We begin with the commutation relation,
\begin{align}
[Q_{K aA}, Q_{D \dot b B}] 
= (\sigma^{\alpha})_{a \dot b} C_{AB} P_{\alpha}\ .
\end{align}
When expanded, this implies
\begin{align}
\left[ Q^{(0)}_K, Q^{(1)}_D \right]=0\ ,
\label{qkqd}
\end{align}
since the kinematical generators $Q_K$ and $P$ have no non-linear parts.

To compute the commutator $\left[ Q^{(0)}_K, Q^{(1)}_D \right]$,
we note that in general,
the commutator between a symmetry generator $\mO$ and the string field
(at the linearised level) is given by
\begin{align}
\left[ \mO^{(0) }, \phi_{P^+}  \right] 
= - \mO \cdot \phi_{P^+} \ .
\label{RFOperatorFieldCommutator}
\end{align}
Here $\mO^{(0)}$ appearing on the LHS denotes
the linear part (quadratic in terms of the fields) of the  charge in the second-quantised
formulation. 
On the RHS $\mO\cdot$ denotes how these operators act on the field
(as a ket-vector) from the left  in the first-quantisation formulation.
The commutator between the charges and $\bphi$ can be computed by
taking the complex conjugate of (\ref{RFOperatorFieldCommutator}).
Apart from the case of a few exceptional operators\footnote{ 
Exceptional ones are $M^{+-}$ and $M^{-\a}$. For these, the ordering 
of $\th$ and $\frac{\d}{\d \th}$ has to be worked out carefully.}, one can show that
\begin{align}
\left[ \mO^{(0) }, \overline{\phi_{P^+}}  \right] 
= \overline{\phi_{P^+}} \cdot \mO \  ,
\label{RFOperatorFieldBarCommutator}
\end{align}
where $\cdot \mO$ denotes the action of the operator
from the right on the complex conjugate of the field (as a bra-vector).
For instance, in the present case,
we have 
\begin{align}
\left[ Q^{(0)}_{KaA}, \phi_{P^+}  \right] 
&= - Q_{KaA} \cdot \phi_{P^+} 
= - \int q_{aA}(\sigma) d\sigma\ \phi_{P^+} \ , \label{eq:rightaction} \\
\left[ Q^{(0)}_{KaA}, \overline{\phi_{P^+}}  \right] 
&= \overline{\phi_{P^+}} \cdot Q_{KaA} 
= - \int d_{aA}(\sigma) d\sigma\  \overline{\phi_{P^+}}\ . \label{eq:leftaction} 
\end{align}  
Since the operator $Q^{(0)}_K$ acts on the string fields,   
\begin{align}
\left[ Q^{(0)}_{K a A}, Q^{(1)}_{D \dot b B} \right] 
&= f^I{}_{JK} 
\int 
\left( 
\overline{\phi_{P^+_3}}_I \cdot Q_{K a A} 
\right)
\left( 
\cdots V     
\right) 
\phi_{P^+_1}{}^J \phi_{P^+_2}{}^K
\prod^3_{r=1} 
dP^+_r \mathcal{D}\theta_r \mathcal{D}x_r \notag \\
&\ \ \  
+ f^I{}_{JK} 
\int 
\overline{\phi_{P^+_3}}_I  
\left( 
\cdots V     
\right) 
Q_{K a A}\cdot \left(\phi_{P^+_1}{}^J \phi_{P^+_2}{}^K \right)
\prod^3_{r=1} 
dP^+_r \mathcal{D}\theta_r \mathcal{D}x_r \ ,
\label{qkqd1}
\end{align}
where
\begin{align}
(\cdots V) =
 (p_{\alpha} \cdot w) (\sigma^{\alpha b}{}_{\dot a} d_{bA} \cdot w) 
(p^+)^{\lambda_0}
(P^+_1)^{\lambda_1}
(P^+_2)^{\lambda_2}
(P^+_3)^{\lambda_3} 
\delta (P^+_1 + P^+_2 - P^+_3) V     
\ . 
\end{align}
Using the associativity property we rewrite (\ref{qkqd1}) as  
\begin{align}
& f^I{}_{JK} 
\int \overline{\phi_{P^+_3}}_I 
\left(
Q^{3}_{KaA} \cdot (\cdots V) + (\cdots V) \cdot  
\left(
Q^{1}_{KaA} + Q^{2}_{KaA}
\right)
\right) 
\phi_{P^+_1}{}^J \phi_{P^+_2}{}^K 
\prod^3_{r=1} 
dP^+_r \mathcal{D}\theta_r \mathcal{D}x_r\ ,
\label{qkqd2}
\end{align}
where $Q^{r}_{K}$ with $r=1,2,3$ 
denotes the operator $Q_K$ acting on the $r$-th string field. 
Moving $Q^{1,2}_K$ to the left of $(\cdots V)$,  
(\ref{qkqd2}) becomes   
\begin{align}
&  f^I{}_{JK} 
\int 
\overline{\phi_{P^+_3}}_I  
\left( 
\int_{I_3}  q_{a A}(\sigma) d\sigma 
+ \int_{I_1}  d_{aA}(\sigma) d\sigma 
+ \int_{I_2}  d_{aA}(\sigma) d\sigma
\right) 
(\cdots V)\
\phi_{P^+_1}{}^J \phi_{P^+_2}{}^K
 \notag \\
& \ \ \ \times 
\prod^3_{r=1}
dP^+_r \mathcal{D}\theta_r \mathcal{D}x_r\ .
\label{qkqd3}
\end{align}
From this we can show  
\begin{align}
&\left( 
\int_{I_3}  q_{a A}(\sigma) d\sigma 
+ \int_{I_1}  d_{aA}(\sigma) d\sigma 
+ \int_{I_2}  d_{aA}(\sigma) d\sigma
\right) 
(\cdots V) \notag \\
&=
\sum_{r=1,2} 
\int_{I_r}
\left[
  d_{aA}(\sigma),  
(\cdots )
\right] d\sigma 
V 
- (\cdots) 
\left( 
\int_{I_3}  q_{a A}(\sigma) d\sigma 
+ \int_{I_1}  d_{aA}(\sigma) d\sigma 
+ \int_{I_2}  d_{aA}(\sigma) d\sigma
\right) V \notag \\
&=0 \ . 
\label{qkqd4}
\end{align}
For the second term in the second line of (\ref{qkqd4}), we used
\begin{equation}
\left( \theta^{ra} (\sigma) - \theta^{3a} (\sigma) \right) V_F = 0\ , \ \ \ 
\left( \frac{\delta}{\delta \theta^{ra} (\sigma) } + \frac{\delta}{\delta \theta^{3a} (\sigma) }  \right) V_F = 0 
\ ,
\end{equation}
where $\sigma \in I_r$ with $r=1,2$, 
and, for the first term, 
\begin{equation}
\int_{I_1} w(\sigma) d\sigma + \int_{I_2} w(\sigma) d\sigma 
=0\ .
\label{RF1dotw}
\end{equation} 
This important property of $w$ is also used  for 
the commutators  
$[M^{+\alpha},Q_D]$ and $[M^{+\alpha},P^{-}_D]$,
which can be verified using similar manipulations.

The commutator $[Q_{D}, M^{\alpha \beta}]$ can also be verified directly.
This is expected since (\ref{RFMAlphaBetaFirstQ}) has the correct index structure ensuring the
correct transformation of $Q_D^{(1)}$ under the SO(4) little group. 

The commutation relation
\begin{equation}
\left[ Q_{D\dot a A}, Q_{D \dot b B} \right] 
= \sqrt{2} \epsilon_{\dot a \dot b} C_{AB} P^- \ ,
\end{equation}
yields 
\begin{equation}
\left[ Q^{(0)}_{D\dot a A}, Q^{(1)}_{D \dot b B} \right] 
+ \left[ Q^{(1)}_{D\dot a A}, Q^{(0)}_{D \dot b B} \right] 
= \sqrt{2} \epsilon_{\dot a \dot b} C_{AB} P^{-(1)} \ .
\end{equation}
Evaluating the LHS, one obtains 
\begin{align}
P^{-(1)} 
&=
2\sqrt{2}\
f^I{}_{JK} 
\int 
\overline{\phi_{P^+_3}}_I 
\left( 
(p_{\alpha} \cdot w)^2  
(p^+)^{\lambda_0}
(P^+_1)^{\lambda_1}
(P^+_2)^{\lambda_2}
(P^+_3)^{\lambda_3} 
\delta (P^+_1 + P^+_2 - P^+_3) V     
\right) \notag \\ 
& \ \ \ \times 
\phi_{P^+_1}{}^J \phi_{P^+_2}{}^K
\prod^3_{r=1} 
dP^+_r \mathcal{D}\theta_r  \mathcal{D}x_r 
\ ,
\label{pminus1}
\end{align}
where we used
\begin{equation}
\left( p^{\alpha}_r (\sigma) + p^{\alpha}_{3}(\sigma) \right) V_B = 0\ ,
\label{RFpVZero}
\end{equation}
for $\sigma \in I_r$ with $r=1,2$. 

Commutators involving $M^{+-}$ can also be verified and lead to the 
same condition (\ref{RFConstrLambdaFromDimAnalysisFinal}) 
obtained from the power counting analysis.
We note that taking the commutator
of the boost generator $M^{+-}$ 
with another operator essentially amounts to counting the number of $P^+$'s
contained in the operator.
One also has to take into account the ``intrinsic weight'', $-\frac12$, 
of the string field under boosts which can be read off from (\ref{RFMPlusMinusFirstQ}).

The commutator $[Q_D , P^-]=0$ requires a careful analysis
using smearing and test functionals, because $p^2$ terms in $P^{-(0)}$
acting on the overlap part, combined with $p\cdot w$ in $Q^{(1)}$ 
may result in unwanted non-zero contributions. 
An outline of this calculation is presented in appendix \ref{RSASmearingTestFunctionalQP}.
The result justifies our choice of the insertion $w$ explained in appendix \ref{RSAInsertion}.

The commutators involving the Lorentz generator $M^{-\a}$ 
are more subtle and we have not completed their analysis. We expect that 
the computation of the commutator 
between $M^{-\a}$ and $P^-$ will fix the $\l$ parameters,
since the analogous parameters of the tensile superstring field theory
were fixed in this way in~\cite{RBLinden}.

\section{Conclusions and discussion}
\label{RSConclusions}

In this paper we have used light-cone string field theory to formulate an 
interacting theory of tensionless strings in six dimensions, with the purpose of
obtaining a Lagrangian description of the $(2,0)$ superconformal field theory. 
Our proposal is motivated by the M-theory picture in which the $(2,0)$ CFT 
arises from the low-energy dynamics of coincident M5-branes. In this M-theory
construction, M2-branes stretched between coincident M5-branes yield 
degrees of freedom consisting of (matrix valued) tensionless closed strings 
confined to the world-volume of the M5-branes. We have argued that string 
field theory, in its light-cone form, is the most suitable language to study these 
interacting tensionless strings. 

The most appealing feature of a formulation of the $(2,0)$ CFT as a 
tensionless string field theory is the fact that it may allow us to avoid the 
obstacles, associated with power counting arguments, which impede the 
construction of local renormalisable interacting QFT's in dimension larger than 
four. The use of stringy degrees of freedom has also interesting implications in 
connection with the relation between the $(2,0)$ CFT in $d=6$ and the 
four-dimensional ${\cal N}=4$ SYM theory. The latter is obtained upon dimensional 
reduction on a torus and we have suggested that wrapped string 
configurations may play a central role in the emergence of the 
four-dimensional Yang--Mills coupling constant. 

In this paper we introduced our formalism and we presented the
free part of the SFT action, together with an ansatz for the cubic interaction
part. These are only the first steps towards obtaining a viable formulation of 
the six-dimensional $(2,0)$ CFT. There remain multiple issues to be clarified,
both of a technical nature -- in the construction of the tensionless SFT -- and 
of a more conceptual nature -- in relation to its interpretation as a description 
of the $(2,0)$ CFT. 

In order to complete the construction of the interacting SFT to cubic order, it 
is important to finish the analysis of the ansatz for the $M^{-\alpha}$ 
Lorentz generators and their commutators with the other charges. We expect
that this should allow us to completely fix our ansatz, determining the 
$\lambda$ parameters. By a more comprehensive study of the constraints 
imposed by the full superconformal algebra, 
one can presumably deduce the full anti-symmetry and
the Jacobi identity for the parameters $f^I{}_{JK}$,
thus characterising them as structure constants of a Lie algebra,
as was done for ${\cal N}=4$ SYM in~\cite{RBABKR}.

Our study of the free part of the superalgebra has only been carried out at the 
level of the Poisson brackets, without taking care of ordering issues in the 
definition of operator products. It is clearly desirable to repeat these 
calculations at the quantum level. For this purpose it may be necessary to 
make a more systematic use of smearing and test functionals, following the 
approach discussed in appendix~\ref{RSASmearingTestFunctional}.

The most important issues that remain to be addressed are, however, more
conceptual and concern the interpretation of our six-dimensional tensionless 
SFT as describing the dynamics of the $(2,0)$ CFT. The fundamental physical 
properties of a CFT formulated in this manner need to be investigated. As a 
theory of tensionless strings our model contains a very large number of light 
degrees of freedom, whose properties and behaviour need to be understood.
The most crucial aspects to focus on are the identification of the correct 
observables in the theory and how to describe them in the SFT language. 
Clarifying these aspects is essential in order to understand the very nature of 
the resulting CFT. 

On general grounds, one expects the proper observables to be correlation
functions of local operators organised in superconformal multiplets. Within the
formulation proposed in this paper such local operators should be built from
the string field. It is possible that there be a vast redundancy in our 
formulation, so that, in spite of the seemingly very large number of degrees of 
freedom contained in the string field, the set of physical observables built from 
them is similar to those found in more familiar conformal theories in lower 
dimensions. It is also possible, however, that the construction that we 
presented give rise to a much broader set of observables compared to more 
conventional CFT's and that the system described by our tensionless string 
field is fundamentally different from the known examples of conformal 
theories. 

There are several ways to gain insights into the properties of observables 
in the theory we constructed. It can be very useful to consider special sectors
in which one has independent means of guessing the structure of the relevant 
observables. Particularly interesting in this respect are large R-charge 
states in M-theory in AdS$_7\times S^4$. According to the AdS/CFT 
correspondence, the $(2,0)$ CFT has a dual description in terms of M-theory 
in AdS$_7\times S^4$, which possesses a large R-charge sector analogous to 
that considered in~\cite{RBKSS1}, described by the BMN matrix model. The 
spectrum of the BMN matrix model includes states associated with near-BPS 
fluctuations of spherical membranes. Then the AdS/CFT duality implies that 
there exist a large R-charge sector in the six-dimensional $(2,0)$ CFT
containing operators corresponding to fluctuations of spherical membranes.
Recalling the properties of the analogous sector in the duality between 
type IIB string theory in AdS$_5\times S^5$ and ${\cal N}=4$ SYM, we can 
speculate about the characteristics of a set of large R-charge degrees of 
freedom in the $(2,0)$ CFT. In the AdS$_5$/CFT$_4$ case one considers 
so-called BMN operators~\cite{RBBMN}, which are constructed as traces 
of products of a large number of the matrix-valued elementary fields of the
${\cal N}=4$ SYM theory. The position in the sequence of fields inside such 
traces can be understood as being associated with the $\s$ coordinate in the 
dual string. In the case of the $(2,0)$ theory the states with large R-charge
we are interested in are membrane fluctuations and thus one has two 
$\s$ coordinates to identify in the relevant CFT operators. 
Since the $(2,0)$ theory contains tensionless string degrees of freedom,
it is natural to build the analog of the BMN operators as traces
of products of matrix fields defined on a loop space, which is the configuration 
space of tensionless strings. In this way one may introduce two 
$\s$-coordinates: one associated with a given ``point'' in the loop space, the 
other labelling the order of the matrix fields in the product. 
Our string field precisely provides a matrix valued field on a loop space.
Thus the
consideration of a BMN-like sector suggests that operators written as traces
of products of string fields 
may be a natural choice of observables in the $(2,0)$ CFT. 
Although in general it is not straightforward to define a theory built on a loop space, 
SFT provides a rather successful example of such a theory. 
This is actually one of the motivations that led us to 
study the SFT approach proposed in this paper. 

When considering the problem of identifying the observables of the $(2,0)$ theory,
it is clearly important to take into account as much as possible the
constraints from symmetry arguments and consistency requirements.
The bootstrap program~\cite{RBBootstrap}
is a way of systematically implement these constraints to obtain, 
in particular, bounds on the spectrum. 
Provided that the observables in our formulation are not of a qualitatively 
different nature from those of more standard CFT's, any constraints 
established using the bootstrap method should be satisfied in our case as 
well.

Further guidance in characterising the observables in our formulation of the 
$(2,0)$ CFT can be provided by the study of the compactification of the theory 
to lower dimensions, in particular to ${\cal N}=4$ SYM in $d=4$. 
Understanding how to derive the ${\cal N}=4$ SYM theory in this way is in its 
own right an important issue, that is essential to address in
order to establish the validity of our formulation. 
Wrapped string configurations are expected to give rise to the SYM degrees
of freedom in $d=4$. However, the fate of unwrapped strings upon 
compactification remains to be clarified. Unless there is a mechanism for the 
decoupling of these configurations, it would appear that our formulation
of the $(2,0)$ CFT may give rise to tensionless strings in four dimensions.
There is also a related issue associated with the presence of an infinite 
number of flat directions (one for each mode of the tensionless string) in the 
action, which may produce severe IR divergences. There seem to be two
possible scenarios in connection to the compactification of our tensionless
SFT to $d=4$ -- either there is a mechanism explaining the decoupling of the 
extra light degrees of freedom or there exists a new description of $\mN=4$ 
SYM in four dimensions containing tensionless strings. 
It would be interesting to study the possible connections of 
such a formulation to the loop equation~\cite{RBMigdalMakeenko, RBMigdal}, 
\ie the Schwinger-Dyson equation for Wilson loop operators, in $\mN=4$ 
SYM. Because of scale invariance, the string arising
from the Wilson loop may be expected to be tensionless.
For our purpose it is natural to consider the loop equation
defined in light-cone superspace~\cite{RBBLN}.
Various types of loop equations for $\mN=4$ SYM,
mainly in the context of the AdS/CFT correspondence, were 
considered in~\cite{RBLoopEqSYM1, RBLoopEqSYM2, RBLoopEqSYM3, 
RBLoopEqSYM4, RBLoopEqSYM5, RBLoopEqSYM6, RBLoopEqSYM7}.

Another important issue to understand is whether or not a critical dimension 
exists for tensionless strings. The analysis of the critical dimension is 
expected to be different compared to the case of ordinary tensile 
strings~\footnote{The Lorentz anomaly in the first-quantised formulation of 
light-cone gauge tensile string theory in six dimensions was computed 
in~\cite{RBMizoguchiKunitomo}. In~\cite{RBLizziRaiSparanoSrivastava, 
RBBonelli} it was argued that there is no critical dimension for tensionless 
bosonic string theory, \ie the theory is consistent for any number of spacetime 
dimensions.}. This is because the nature of the UV divergences in 
$\s$-space and the normal ordering, which underlie the calculation of the
critical dimension, are different in the tensionless case.
Moreover, in the case we are interested in the coupling constant should be of 
order $1$ and thus the free and the interaction parts may mix 
when discussing possible anomalies in the Lorentz symmetry.

The possible mixing between contributions of different orders has another 
important implication. It may allow us to determine the magnitude of the 
coupling constant by requiring the cancellation of the quantum anomaly in 
the symmetry algebra. In the case of the bosonic open-closed light-cone 
gauge string field theory, it is known that the Lorentz anomaly of the string 
field theory (not that of the first quantised theory) determines the relationship 
between the various coupling constants in the theory~\cite{RBSaitohTanii1, 
RBSaitohTanii2, RBKikkawaSawada}.
The situation in the case of the $(2,0)$ CFT may be analogous to that of 
the Chern-Simons theory, in which the coupling constant is constrained to be 
an integer by the requirement that the path integral be uniquely defined.
Another way to fix the coupling constant is to work out the
reduction discussed above to four-dimensional $\mN=4$ SYM.

Our formulation of the (2,0) theory as a tensionless string theory for the 
low-energy dynamics of M5-branes is analogous to the description of 
the low-energy dynamics of parallel D-branes in terms of SYM theories.
In view of this, we expect to have the analog of the
well-known realisation of the Higgs mechanism in a system of D-branes.
Each matrix element of our matrix-valued string field contains
the tensor multiplet arising from the zero mode part of $x(\s)$ 
and $\th(\s)$. The 5 scalar fields in the tensor multiplet describe 
transverse fluctuations of the M5-branes and a vacuum expectation value 
for the scalars in the $i$-th diagonal element in the matrix-valued string field
corresponds to the position of the $i$-th M5-brane.
It is interesting to study the theory around configurations in which these
scalar fields have non-zero vacuum expectation values.
The theory should then describe the low energy limit of parallel,
but non-coincident, M5-branes. There are two scales involved in this
construction, the M2-brane tension (or equivalently the 11-dimensional 
Planck length) and the separation between the M5-branes (or equivalently 
the scalar vacuum expectation value). The tension of the strings arising 
from M2-branes stretched between M5-branes is the product of the 
membrane tension and the distance between the M5-branes. One should 
consider the low energy limit by simultaneously sending to zero the 
separation between any two M5-branes, in such a way as to keep the 
tension of the strings finite when measured in terms of 
the relevant energy scale. Equivalently, one sends the eleven-dimensional 
Planck energy to infinity, while tuning the distances between M5-branes, so 
that the string tension remains finite.
Let us consider, for definiteness, the case in which $N$ 
M5-branes are divided into two groups of $N_1$ and $N_2$ coincident 
branes, with $N=N_1+N_2$. The configuration is then represented by
a block diagonal matrix. In the original $N\times N$ matrix one can identify 
$N_1 \times N_1$ and $N_2 \times N_2$ diagonal blocks  
and two off-diagonal blocks of size $N_1 \times N_2$
and $N_2 \times N_1$ respectively.
The scaling limit should decouple both the bulk gravity dynamics and the
degrees of freedom associated with fluctuations of the M2-branes in the directions transverse to the M5-branes.
In this limit the DOF contained in the block diagonal elements
should be tensionless strings
and those contained in the block off-diagonal elements should be
tensile strings with a tension proportional to the vacuum expectation value
(or equivalently the distance between the two sets of M5-branes).
This coupled system of tensionless and tensile strings should arise 
by expanding our SFT around the configuration with 
non-zero vacuum expectation values. In this situation the cubic and higher 
order vertices in the Hamiltonian give rise to additional contributions to the
part quadratic in the string fields. Checking that these quadratic terms  
produce the correct free Hamiltonian for the block off-diagonal tensile strings
provides a non-trivial test of the form of the interaction vertices.

One may also study M5-branes in a spacetime with a compactified transverse 
direction, that can be realised considering an infinite number of copies of 
M5-branes, in a way analogous to the description of D-branes in a 
compactified spacetime by SYM~\cite{RBTaylor}.
In this way one may obtain a SFT formulation of the theory 
describing the decoupling limit of NS5-branes, \ie
the little string theory with (2,0) supersymmetry~\cite{RBSeibergLittleString}.
For a review of little string theory, see \cite{RBAharony}. 
The SFT description would contain tensionless strings as well as an infinite 
variety of tensile strings with tensions proportional to an integer multiple
of the compactification radius.

In this paper we constructed the cubic vertex for a tensionless string
field theory in six dimensions. It is important to study the possible higher 
order terms in the Hamiltonian.  
In the case of tensile bosonic string field theory in light-cone gauge, it is 
known that cubic and quartic vertices are sufficient to reproduce the
correct S-matrix~\cite{RBKakuKikkawa1, RBKakuKikkawa2}. 
For the light-cone superstring field theory constructed 
in~\cite{RBGS1,RBGSB,RBGS2,RBGS3} the necessity of quartic 
couplings was discussed in~\cite{RBGreenSeiberg, RBGreensiteKlinkhamer1, 
RBGreensiteKlinkhamer2, RBGreensiteKlinkhamer3}, but there seems to
be no definitive answer to the question of whether higher order vertices 
are present in the theory. 
It is still premature to draw any conclusions about the structure of 
higher-order terms in our model, although the
similarity and close relationship to $\mN=4$ SYM may suggest that the  
action should stop at quartic order.

The SFT description we proposed in this paper 
applies to a special sector of M-theory, \ie
the low energy fluctuations of coincident M5-branes.
The tensionless string DOF we studied arise from
membranes stretched between coincident M5-branes. 
The matrix model of M-theory~\cite{RBdeWitHoppeNicolai,RBBFSS},
which is a good candidate for the formulation of the full M-theory,
can be considered as the matrix-regularised version of 
membrane theory~\cite{RBGoldstone, RBHoppe, RBdeWitHoppeNicolai}. 
Within this framework it is possible that our SFT construction
may eventually be superseded by a description in terms of regularised DOF.

Although additional work is required to establish whether our tensionless 
string field theory approach will lead to a valid formulation of the  
six-dimensional $(2,0)$ CFT, we believe that the ideas presented in this paper 
deserve to be further studied. If successful, this proposal would extend the 
realms of both string theory and QFT. We hope that our work provides the first 
steps and the necessary tools to pursue this line of investigation.

\vspace*{0.7cm}
\ndt
{\bf Acknowledgements}

\vspace*{0.3cm}
\ndt
We are grateful to Pierre Ramond for sharing unpublished results with us.
We thank Lars Brink and Pierre Ramond for discussions. 
Part of this work was done when HS was at the Okayama Institute for Quantum Physics and
at KEK. HS is grateful to his colleagues there in particular Fumihiko Sugino, Satoshi Iso
and Shotaro Shiba for many useful discussions, comments and encouraging remarks.
We would also like to thank 
Koji Hashimoto, Nobuyuki Ishibashi, Hiroshi Isono,
Hikaru Kawai, Yoichi Kazama, Seok Kim, Ryuichiro Kitano, Shota Komatsu, Hiroshi Kunitomo, Tsunehide Kuroki, 
Shun'ya Mizoguchi,
Norisuke Sakai, Tadakatsu Sakai, Ashoke Sen, Shigeki Sugimoto, 
Stefan Theisen, Seiji Terashima,
Satoshi Yamaguchi and Tamiaki Yoneya for discussions, encouraging remarks and useful comments. 
The work of YS was supported by Building of Consortia for the Development of Human Resources 
in Science and Technology and by CUniverse research promotion project by Chulalongkorn University 
(grant reference CUAASC).  
The work of HS was supported by JSPS KAKENHI Grant Numbers JP16H06490.
The work of SA is partially supported by a DST-SERB grant
(EMR/2014/000687).

\appendix

\numberwithin{equation}{section}

\section{Tensors with R-symmetry and spinor indices}
\label{RSATensors}
Six-dimensional $\mN=(2,0)$ supersymmetry is described, for example,
in \cite{RBHoweSierraTownsend, RBStrathdee}. 

\subsection{R-symmetry USp(4)}
For the R-symmetry USp(4) tensors we use an anti-symmetric and 
non-degenerate $4\times 4$ matrix $C$,
\begin{align}
C_{AB}=-C_{BA}\ .
\end{align}
It is related to the $B$ matrix used in the complex conjugation by
\begin{align}
C=B^TA\ ,
\end{align}
\ie
\begin{align}
C_{AB} = B^{\bC}{}_{A} A_{\bC B}\ ,
\end{align}
where one can choose a representation in which $A$ equals the Kronecker delta.
The $B$ matrix satisfies
\begin{align}
B^* B=-1\ ,
\end{align}
\ie
\begin{align}
B^{\bA}{}_B \overline{B^{\bB}{}_C} =- \d^{\bA}{}_{\bC}\ .
\end{align}

\subsection{Light-cone little group SO(4)}
We define SU(2) anti-symmetric $\epsilon$ tensors with the convention
\begin{align}
\e^{12}=&1\ ,
\\
\e_{12}=&1\ .
\end{align}
We introduce the $\s$-matrices
\begin{equation}
(\sigma^{\alpha})^{\dot a b} = - (\sigma^{\alpha})^{b \dot a}\ , \ \ \ 
(\sigma^{\alpha})_{\dot a b} = - (\sigma^{\alpha})_{b \dot a}\ ,
\end{equation}
related to each other by
\begin{align}
\s^{\a a \db} =&+\e^{ac} \e^{\db\dd} \s^\a_{c \dd}\ ,
\\\s^\a_{c\dd}
=& 
\s^{\a a \db} \e_{ac} \e_{\db \dd}\ .
\end{align}
They satisfy the algebra
\begin{align}
\s^{\a a \dc} 
\s^{\b}{}_{ \dc b} 
+
\s^{\b a \dc} 
\s^{\a}{}_{\dc b} 
=&\d^a{}_b\ ,
\\
\s^{\a \da c} 
\s^{\b}{}_{ c \db} 
+
\s^{\b \da c} 
\s^{\a}{}_{c \db} 
=&\d^\da{}_\db\ .
\end{align}
An explicit representation is
\begin{align}
\s^\a{}^{a \db}
&=(-\s^1, \s^2, -\s^3, i 1)\ ,
\\
\s^\a{}^{\da b}
&=(\s^1, \s^2, \s^3, -i 1)\ .
\end{align}

We define
\begin{align}
\s^{\a\b}{}^a{}_b
&=\frac{1}{2}\left(
\s^{\a a \dc} \s^{\b}{}_{\dc b}
-
\s^{\b a \dc} \s^{\a}{}_{\dc b}
\right)\ ,
\\
\s^{\a\b}{}^{\da}{}_{\db}
&=\frac{1}{2}\left(
\s^{\a \da c} \s^{\b}{}_{c \db}
-
\s^{\b \da c} \s^{\a}{}_{c \db}
\right)\ 
\end{align}
and
\begin{align}
\s^{\a\b a b}
=
\s^{\a\b a}{}_c \e^{cb}\ ,
\end{align}
which satisfy
\begin{align}
\s^{\a\b\, cd}
=\s^{\a\b\, dc}\ .
\end{align}
We introduce $2\times 2$ matrices 
$B^{\bda}{}_\dc$,
$B^{\ba}{}_c$,
whose components are equal to those of 
$-i \s^2$.
We have
\begin{align}
B^{-1\, c}{}_{\ba} 
\overline{
\s^{\a a \db}}
B^{-1\, \dd}{}_{\bdb} 
=&
+ 
\s^{\a c \dd}\ ,
\\
B^{\bda}{}_{\dc} 
\overline{
\s^{\a}{}_{\da b}}
B^{\bb}{}_{d} 
=&
+
\s^{\a}{}_{\dc d}\ ,
\\
B^{-1\, c}{}_{\ba} 
\overline{
\s^{\a\b\, a}{}_b}
B^{\bb}{}_{d} 
=&
+\s^{\a\b\, c}{}_d\ ,
\\
\overline{\e^{cd}} B^{-1\, e}{}_\bd =&
+B^{\bc}{}_d \e^{de}\ ,
\\
B^{-1\, d}{}_{\bb} 
B^{-1\, c}{}_{\ba} 
\overline{
\s^{\a\b\, ab}}
=&
\s^{\a\b\, cd}\ .
\end{align}

\section{Superalgebra}  
\label{RSASuperAlgebra}
\vskip 0.5cm

\begin{align}
&[ (Q_K)_{aA} , (Q_K)_{bB} ] = - \sqrt{2} \epsilon_{ab} C_{AB}P^+\ , \\
&[ (Q_K)_{aA} , (Q_D)_{\dot b B} ] = (\sigma^{\alpha})_{a \dot b} C_{AB} P_{\alpha}\ , \\
&[ (Q_D)_{\dot a A}, (Q_K)_{bB} ] = - (\sigma^{\alpha})_{b \dot a} C_{AB} P_{\alpha}\ , \\
&[ (Q_D)_{\dot a A}, (Q_D)_{\dot b B}] = \sqrt{2} \epsilon_{\dot a \dot b} C_{AB} P^-\ , \\
&[ M^{+ \alpha}, (Q_{D})_{\dot a A}] = - \frac{i}{\sqrt{2}} (Q_K)_{bA} \epsilon^{bc} (\sigma^{\alpha})_{c \dot a}\ , \\
&[ M^{-\alpha}, (Q_K)_{aA} ] = \frac{i}{\sqrt{2}} (Q_D)_{\dot b A} \epsilon^{\dot b \dot c} (\sigma^{\alpha})_{\dot c a}\ , \\
&[ M^{\alpha \beta}, (Q_K)_{aA} ] = - \frac{i}{2} (Q_K)_{bA} (\sigma^{\alpha \beta})^b{}_{a}\ , \\
&[ M^{\alpha \beta}, (Q_D)_{\dot a A} ] = - \frac{i}{2} (Q_D)_{\dot b A} (\sigma^{\alpha \beta})^{\dot b}{}_{\dot a}\ , \\
&[ M^{+-}, (Q_K)_{aA} ] = \frac{i}{2} (Q_K)_{aA}\ , \\
&[ M^{+-}, (Q_D)_{\dot a A} ] = - \frac{i}{2} (Q_D)_{\dot a A}\ , \\
&[ M^{+-}, M^{+\alpha} ] = i M^{+\alpha}\ , \\
&[ M^{+-}, M^{-\alpha} ] = -i M^{-\alpha}\ , \\
&[ M^{+\alpha}, M^{-\beta} ] = -iM^{\alpha \beta} + i \delta^{\alpha \beta} M^{+-}\ , \\
&[ M^{\alpha \beta}, M^{\pm \gamma} ] = i ( \eta^{\alpha \gamma} M^{\pm \beta} - \eta^{\beta \gamma} M^{\pm \alpha} )\ , \\
&[ M^{\alpha \beta}, M^{\gamma \delta} ] = i ( \eta^{\beta \gamma}M^{\delta \alpha} - \eta^{\alpha \gamma} M^{\delta \beta} - \eta^{\beta \delta} M^{\gamma \alpha} + \eta^{\alpha \delta} M^{\gamma \beta} )\ , \\
&[ M^{+-},P^+ ] = iP^+\ , \\
&[ M^{+-}, P^- ] = - i P^-\ , \\
&[ M^{+\alpha}, P^- ] = - i P^{\alpha}\ , \\
&[ M^{+\alpha}, P^{\beta} ] = -iP^+ \delta^{\alpha \beta}\ , \\
&[ M^{-\alpha}, P^+ ] = -i P^{\alpha}\ , \\
&[ M^{-\alpha}, P^{\beta} ] = - i P^- \delta^{\alpha \beta}\ , \\
&[ M^{\alpha \beta}, P^{\gamma} ] = i (P^{\beta} \delta^{\gamma \alpha} - P^{\alpha} \delta^{\gamma \beta})\ .
\end{align}
All other commutators not listed here vanish. 

Our convention is that all bosonic charges $M$'s and $P$'s are hermitian, 
while $Q_K$ and $Q_D$ satisfy the hermiticity conditions
\begin{align}
\overline{Q_{KaA}} 
&= - Q_{KbB}\ B^{-1}{}^{b}{}_{\bar a} B^{-1}{}^B{}_{\bar A}\ ,
\label{Qkhermiticity} \\
\overline{Q_{D\dot{a} A}} 
&= Q_{D \dot{b} B}\ B^{-1}{}^{\dot{b}}{}_{\bar{\dot{a}}} B^{-1}{}^B{}_{\overline A}\ .
\label{Qdhermiticity}
\end{align}

\section{Computation of $[M^{-\alpha},M^{-\beta}]$}  
\label{RSAMM}
\vskip 0.5cm

We verified explicitly the commutators between the charges for the free part of the theory
presented in section \ref{RSFree}. We work at the level of Poisson brackets,  
\ie we ignore ordering issues in the definition of 
products of operators.

In this appendix we show how to compute the commutators 
of the free part of the symmetry charges
focussing on the most involved commutator 
\begin{equation}
[M^{-\a}, M^{-\b}] = 0\ ,
\end{equation}
as an example.

For the free part, we can work solely in the first quantised language,
\begin{align}
M^{-\a} 
&= \int_0^{[\s]} \Bigg( x^- (\s) p^{\a}(\s) 
- x^{\a}(\s) p^- (\s) \notag \\
& \ \ \ 
+\frac{i}{2} \theta^{aA}(\s) \frac{\delta}{\delta \theta^{aA}} (\s) \frac{p^{\a}(\s)}{p^+} 
+ \frac{\sqrt{2}}{8} i \frac{p^{\gamma} (\s)}{(p^+)^2} q_{aA} (\s) q_{bB} (\s) \s^{\a \gamma}{}^{ab} C^{-1 AB}\Bigg)d\s\ .
\end{align}
For simplicity we choose the range of $\s$ to be $[0,[\s]]$;
the computation goes through also in the convention used in the main text.

The essential simplification which occurs for the tensionless string theory
is that a good part of the computation is completely parallel to
the computation for the superparticle case. 
This is because each charge presented in section \ref{RSFree} is 
an integral of the charge density which does not involve $\s$-derivatives.
Dropping the $\s$ dependence from the charge density, we get the charge 
for the superparticle case.
Thus for example $M^{-\a}$ for the superparticle is
\begin{align}
M^{-\a}= x^-  p^{\a} 
- x^{\a} p^-  
+\frac{i}{2} \theta^{aA} \frac{\der}{\der \theta^{aA}}  \frac{p^{\a}}{p^+} 
+ \frac{\sqrt{2}}{8} 
i \frac{p^{\gamma} }{(p^+)^2} q_{aA}  q_{bB}  \s^{\a \gamma}{}^{ab} C^{-1 AB}\ .
\end{align}
The definition of $q$ is the same as (\ref{RFDefq}) except that there is no $\s$-dependence 
for the superparticle case.
By a slight abuse of notation, we use for the variables characterising the  
superparticle, $x^+, p^-, x^\a, p_\a, \th^{aA}$, the same symbols used in the 
string case. The commutation relations between these variables are 
\begin{align}
[x^+, p^-]=&-i\ ,
\\
[x^\a, p^\b]=&i \d^{\a\b}\ ,
\\
\left[ \frac{\der}{\der \th^{aA}}, \th^{bB} \right]
=&\d^a{}_b \d^A{}_B\ .
\end{align}
Comparing these to the commutation relations in the tensionless superstring theory
\begin{align}
[X^+, P^-]=& -i\ , 
\\
[x^\a(\s), p^\b(\s')]=&i \d^{\a\b} \d(\s-\s')\ ,
\label{RFOtherCommutatorApp1}
\\
[\frac{\d}{\d \th^{aA}}(\s), \th^{bB}(\s')]=&\d^a{}_b \d^A{}_B \d(\s-\s')\ ,
\label{RFOtherCommutatorApp2}
\end{align}
we see that if $x^-(\s)$ is not involved, 
the computation of commutators for the tensionless string case
is completely parallel to the superparticle case; the commutators
between the charge densities of the tensionless string are given simply by
the commutators between the charges of the particle multiplied by $\d(\s-\s')$.

The only charge~\footnote{$M^{+-}$ depends only 
on the zero-mode $X^-$.} which contains $x^-(\s)$ is $M^{-\a}$.
Hence one needs to perform additional computations to verify the 
commutation relations involving this generator.
In section \ref{RSAMMParticle} we present the computation of the commutator
$[M^{-\a}, M^{-\b}]$ in the superparticle case and in \ref{RSAMMXMinus}
we explain the modifications necessary to deal with the tensionless 
superstring case.

\subsection{Superparticle case} 
\label{RSAMMParticle}
We write the generator as
\begin{equation}
M^{-\a}  
= X^{\a} + Y^{\a}\ , 
\end{equation}
where
\begin{equation}
X^{\a} = x^-  p^{\a}  
- x^{\a} p^- 
+\frac{i}{2} \theta^{aA} \frac{\der}{\der \theta^{aA}}  \frac{p^{\a}}{p^+}\ , 
\ \ \ 
Y^{\a} 
= 
\frac{\sqrt{2}}{8} i \frac{p^{\gamma} }{(p^+)^2} q_{aA} q_{bB} \s^{\a \gamma}{}^{ab} C^{-1 AB}\ . 
\end{equation}
It is easy to show 
\begin{align}
[X^\a, X^\b] &= 0\ , \\ 
[X^\a, Y^\b]
&=
- \frac{\sqrt{2}}{8} 
\frac{p^\g p^\a}{(p^+)^3}
q_{aA}
q_{bB}
\s^{\b\g}{}^{ab}
C^{-1 AB}
-\frac{\sqrt2}{8}\frac{p^-}{(p^+)^2}
q_{aA} q_{bB} \s^{\a\b a b} C^{-1 AB}\ .
\end{align}
We also get  
\begin{align}
[Y^\a, Y^\b] 
&=
\left(\frac{\sqrt2}{8} i\right)^2\times 2\times 
\frac{p^\g}{(p^+)^2}
\frac{p^\d}{(p^+)^2}
\times
\s^{\ul{\a}\g ab} C^{-1 LM}
\s^{\ul{\b}\d cd} C^{-1 NP}
\times
q_{aL}
[q_{bM}, 
q_{cN}]
q_{dP}
\notag 
\\
&=
\frac{1}{16} 
\frac{p^\g p^\d}{(p^+)^4}
\times
\s^{\ul{\a}\g ab} C^{-1 LM}
\s^{\ul{\b}\d cd} C^{-1 NP}
\times
q_{aL}
q_{dP}
\times
\sqrt{2}\e_{bc} C_{MN} p^+
\notag 
\\
&=
-
\frac{\sqrt2}{8} 
\frac{1}{(p^+)^3}
\times
q_{aL}
q_{dP}
\times
C^{-1 LP}
\times
\left(
\s^{\ul{\a}\g ad} p_\g p^{\ul{\b}}
-
p_\g p^\g \s^{\a \b a d}
\right)\ ,
\end{align}
where the underlined indices are understood to be anti-symmetrised with
no multiplicative coefficient. 
Adding up these contributions, we obtain $[M^{-\a}, M^{-\b}]=0$ for the superparticle case.

In the computation we use the following formulae and the general formulae listed in 
appendix \ref{RSATensors} 
\begin{align}
&\s^{\ul{\a}\g ab}
\e_{bc}
\s^{\ul{\b}\d cd}
=
-
\s^{\ul{\a}\g a}{}_{b}
\s^{\ul{\b}\d b}{}_{c}
\e^{cd}\ ,
\\
&\s^{\a\g}
\s^{\b\d}
=
\s^{\a\g\b\d}
+
\s^{\a\d}\d^{\g\b}
-
\s^{\g\d}\d^{\a\b}
-
\s^{\a\b}\d^{\g\d}
+
\s^{\g\b}\d^{\a\d}
+
\d^{\a\d}\d^{\g\b}-\d^{\a\b}\d^{\g\d}\ , \\
&\s^{\ul{\a}\g}
\s^{\ul{\b}\d}
=
2 \s^{\a\g\b\d}
+
\s^{\a\d}\d^{\g\b}
-2
\s^{\a\b}\d^{\g\d}
+
\s^{\g\b}\d^{\a\d}
-
\s^{\b\d}\d^{\g\a}
-
\s^{\g\a}\d^{\b\d} 
+
\d^{\a\d}\d^{\g\b}-\d^{\b\d}\d^{\g\a}\ , \\
&\s^{\ul{\a}\g}
\s^{\ul{\b}\d}
p_\g p_\d
=
2\s^{\a\g} p_{\g} p^\b
-2\s^{\a\b} p_{\g} p^\g
-2\s^{\b\g} p_{\g} p^\a\ .
\end{align}
In the last three equations the spinor indices are suppressed.

\subsection{Contribution involving $x^-(\s)$ in $[M^{-\a}, M^{-\b}]$}
\label{RSAMMXMinus}
As already explained, most of the terms appearing 
in the computation of $[M^{-\a}, M^{-\b}]$ for the tensionless superstring case
can be simply obtained from the corresponding terms in the computation for 
the superparticle.

The exceptions are the terms involving $x^-$, since $x^-$ is defined non-locally
in terms of other dynamical variables (\ref{RFDefXMinus}). 
More practically, the calculations in the string and in the particle cases
differ because $p^+$ is a c-number in the string case 
and we do not have the analogue of the commutator
\begin{align}
[ x^-(\s) , p^+(\s')] = -i \d(\s-\s')\ .
\label{RFcomofxmpp}
\end{align}

The term involving $x^-$ in the $[M^{-\alpha},M^{-\beta}]$ commutator is
\begin{equation}
[A^{-\alpha}, M^{-\b}]\ , \ \ \ 
\text{with}\ \ \ 
A^{-\a} 
= \int^{[\s]}_0 x^-(\s') p^{\alpha} (\s')\ d\s'\ .
\label{RFDefAMinusAlpha}
\end{equation}

This commutator can be computed by rewriting
the generator $A^{-\a}$ following Mandelstam \cite{RBMandelstamLorentz},   
\begin{align}
A^{-\a} 
&=X^- P^{\alpha}
+ \int^{[\s]}_0 x^- (\s') 
\left( 
p^{\alpha} (\s') - \frac{P^{\alpha}}{[\s]} 
\right)d\s' 
\notag \\
&= X^- P^{\alpha} 
+ \int^{[\s]}_0  
x^-(\s')\ \partial_{\s'} 
\int^{\sigma'}_0 
\left( 
p^{\alpha} (\s'') - \frac{P^{\alpha}}{[\s]} 
\right)d\s'' d\s' \notag \\
&= X^- P^{\a}
- \int^{[\s]}_0  
\partial_{\s'}x^-(\sigma') 
\left(
\int^{\s'}_0 p^{\alpha}(\s'')\ d\s'' 
- \frac{P^{\a}}{[\s]} 
\s' 
\right)d\s'\ , 
\label{RFMTrick}
\end{align}
where $\partial_{\s}x^-$ is given by (\ref{RFDefXMinus}). 

The computation of $[A^{-\a}, M^{-\b}]$ can be done systematically 
by noting the following observation about
the commutator $[A^{-\a}, f]$ for a generic dynamical variable $f$. 
We denote 
by $[A^{-\a}, f]_{\text{cov}}$  
the commutator based on the covariant commutation relation, \ie
the commutation relations (\ref{RFcomofxmpp}), (\ref{RFOtherCommutatorApp1})
and (\ref{RFOtherCommutatorApp2}).
The computation of $[A^{-\a}, f]_{\text{cov}}$ can be done in a way which 
is completely parallel to the superparticle case.
Thus the difference between 
$[A^{-\a}, f]$ and $[A^{-\a}, f]_{\text{cov}}$
is of interest.

It is known that this difference can be understood
as the effect of the compensating gauge transformation on $f$~\cite{RBGGRT}.
The generator $A^{-\a}$ (using the covariant commutation relation) 
transforms $p^+$. This breaks the light-cone gauge condition and 
one needs a compensating gauge transformation to go back to 
the light-cone gauge slice.  
The variation of $p^+$ computed using the covariant commutation relation is    
proportional to
\begin{equation}
[A^{-\a}, p^+]_{\text{cov}} 
= - i p^{\a} (\s)\ . 
\label{dpplus1}
\end{equation}
Since $p^+$ transforms as a density under $\s$-reparametrisations, we have  
\begin{equation}
\delta p^+ (\s) 
= - \partial_{\s} (p^+ \delta \s (\s)) 
= -p^+ \partial_{\s} \delta \s (\s)\ .
\label{dpplus2} 
\end{equation} 
Comparing (\ref{dpplus1}) with (\ref{dpplus2}) we find 
that $\delta \s$ associated with the compensating gauge transformation
is proportional to 
\begin{equation}
u^\a (\s) 
= \frac{1}{p^+} \int^{\s}_{0} p^{\alpha} (\s')\ d\s' 
\ ,
\label{RFDSigma}
\end{equation} 
where the integration constant is fixed by $\delta \s (0) = 0$.
We have
\begin{align}
[A^{-\a}, f(\s)] &= 
[A^{-\a}, f(\s)]_{\text{cov}}
+ i \partial_{\s} f(\s) u^\a (\s) 
\ \ \  \text{if $f$ is a scalar}\ , \label{RFComOfAFs} \\ 
[A^{-\a}, f(\s)] &= 
[A^{-\a}, f(\s)]_{\text{cov}}
 + i \partial_{\s} ( f(\s) u^\a (\s) )  
\ \ \ \text{if $f$ is a density}\ . \label{RFComOfAFd}
\end{align}
The second terms on the RHS correspond to the compensating gauge transformations. 
Indeed, for $f=p^+$ the RHS of (\ref{RFComOfAFd}) vanishes~\footnote{
There are two conventions for the light-cone gauge in string theory.
The convention we are using in which $p^+$ is a constant 
is suitable when discussing 
interactions of strings~\cite{RBMandelstamCubicVertex}.
There is another convention,
used in~\cite{RBGGRT},
in which $[\s]$ is a constant (such as $2\pi$).
The form of the compensating gauge transformation depends on this convention.
In the convention of \cite{RBGGRT}, we need 
another contribution to the RHS of (\ref{RFDSigma}) which is linear in $\s$.}.

Later, we will need to evaluate 
$\left[ A^{-\a}, \int^{[\s]}_0 f(\s)\ d\s
\right]$.
We have
\begin{equation}
\left[
A^{-\a}, \int^{[\s]}_0 f(\s)\ d\s
\right] 
= - i \frac{P^{\a}}{p^+} f([\s]) 
+ \int^{[\s]}_0 [A^{-\a}, f(\s)]\ d\s
\ . 
\label{amaintf}
\end{equation}
To obtain the first term, we regularise the integral in terms of a   
Riemann sum, 
\begin{equation}
\int^{[\s]}_0 f(\s)\ d\s 
\cong 
 \sum^{M}_{m=1} f \left( \frac{[\s]}{M} m \right) 
\frac{[\s]}{M}
\ ,
\end{equation}
and use
\begin{equation}
[X^-, [\s]] 
= \left[ X^-, \frac{P^+}{p^+} \right]
= - \frac{i}{p^+}\ .
\end{equation} 
In particular, if $f$ is a density, we obtain 
\begin{equation}
\left[
A^{-\a}, \int^{[\s]}_0 f(\s)\ d\s
\right] =  \int^{[\s]}_0 [A^{-\a}, f(\s)]_{\text{cov}} \ d\s\ .
\label{RFAmaIntF}
\end{equation}
using (\ref{RFComOfAFd}).

We compute $[A^{-\a}, M^{-\b}]$ by 
successively verifying the generic formulae
(\ref{RFComOfAFs}) and (\ref{RFComOfAFd}) for various building blocks of $M^{-\b}$.
For instance, we verify (\ref{RFComOfAFs}) for $f=x^-$, 
and (\ref{RFComOfAFd}) for $f=p^\b$, and then 
(\ref{RFComOfAFd}) for $f=x^-p^\b$. 
Finally by using (\ref{RFAmaIntF})  
we obtain  
\begin{equation}
[A^{-\a}, M^{-\b}] 
= 
[A^{-\a}, M^{-\b} ]_{\text{cov}}\ .
\end{equation}
This, combined with the computation for the superparticle
in appendix \ref{RSAMMParticle}, implies $[M^{-\a}, M^{-\b}] = 0$.

The following formulae are used in the computation.  
$x^-(\s)$ can be written as~\cite{RBGGRT}   
\begin{equation}
x^-(\sigma) 
= X^- 
+ \int^{[\sigma]}_0 
\left(
\frac{\sigma'}{[\sigma]} 
- \theta (\sigma' - \sigma)
\right) \frac{1}{p^+}
\left(
p^{\beta} (\sigma') \partial_{\sigma}x^{\beta} (\sigma') 
- i \partial_{\sigma} \theta^{aA} (\sigma') \frac{\delta}{\delta \theta^{aA}} (\sigma')
\right) d \sigma'\ , 
\label{RFXm}
\end{equation}
which can be confirmed by differentiating with respect to $\s$ and using (\ref{RFDefXMinus}). 
When computing $[A^{-\a},x^-(\s)]$ 
the integral over $\s$ in (\ref{RFXm}) should be dealt with in a manner similar 
to the manipulations used above for the computation of  (\ref{amaintf}).
Another important formula is
\begin{align}
[X^-,x^\b(\s)] 
&=\left[ X^-, \sum_n x^{\b n} e^{in\frac{2\pi}{[\s]}\s} \right] 
= i \partial_\s x^\b (\s) \frac{\s}{P^+}\ . \label{RFXmx} 
\end{align} 
We also use 
\begin{align} 
[X^-,p^\b(\s)] &= i \partial_\s \left( p^\b (\s) \frac{\s}{P^+} \right)\ , \label{RFXmp} \\
[X^-, \theta^{aA}(\s)] &= i\partial_\s \theta^{aA}(\s)  \frac{\s}{P^+}\ ,\label{RFXmtheta} \\  
\left[ X^- , \frac{\delta}{\delta \theta^{aA}}(\s) \right] &= i \partial_\s \left( \frac{\delta}{\delta \theta^{aA}}(\s) \frac{\s}{P^+} \right)\ . \label{RFXmdeltheta}
\end{align}

\section{Overlap and insertion}
\label{RSAOverlapInsertion}
\subsection{Insertion operator}
\label{RSAInsertion}
In this appendix we motivate the use of $w(\s)$ defined in the main text
(\ref{RFDefwFirst})-(\ref{RFDefwLast})
as the insertion and we discuss an alternative possibility.

One should insert operators at the interaction point, since there is no other
special point on the string world-sheet.
It is necessary here to distinguish the immediate left/right of the interaction 
point, since the very concept of interaction point may be considered as defined by 
the change of left/right from the point of view of 
the $r=1,2$ strings and the $r=3$ string.

\begin{figure}[h]
\centering
\includegraphics[width=0.6\textwidth]{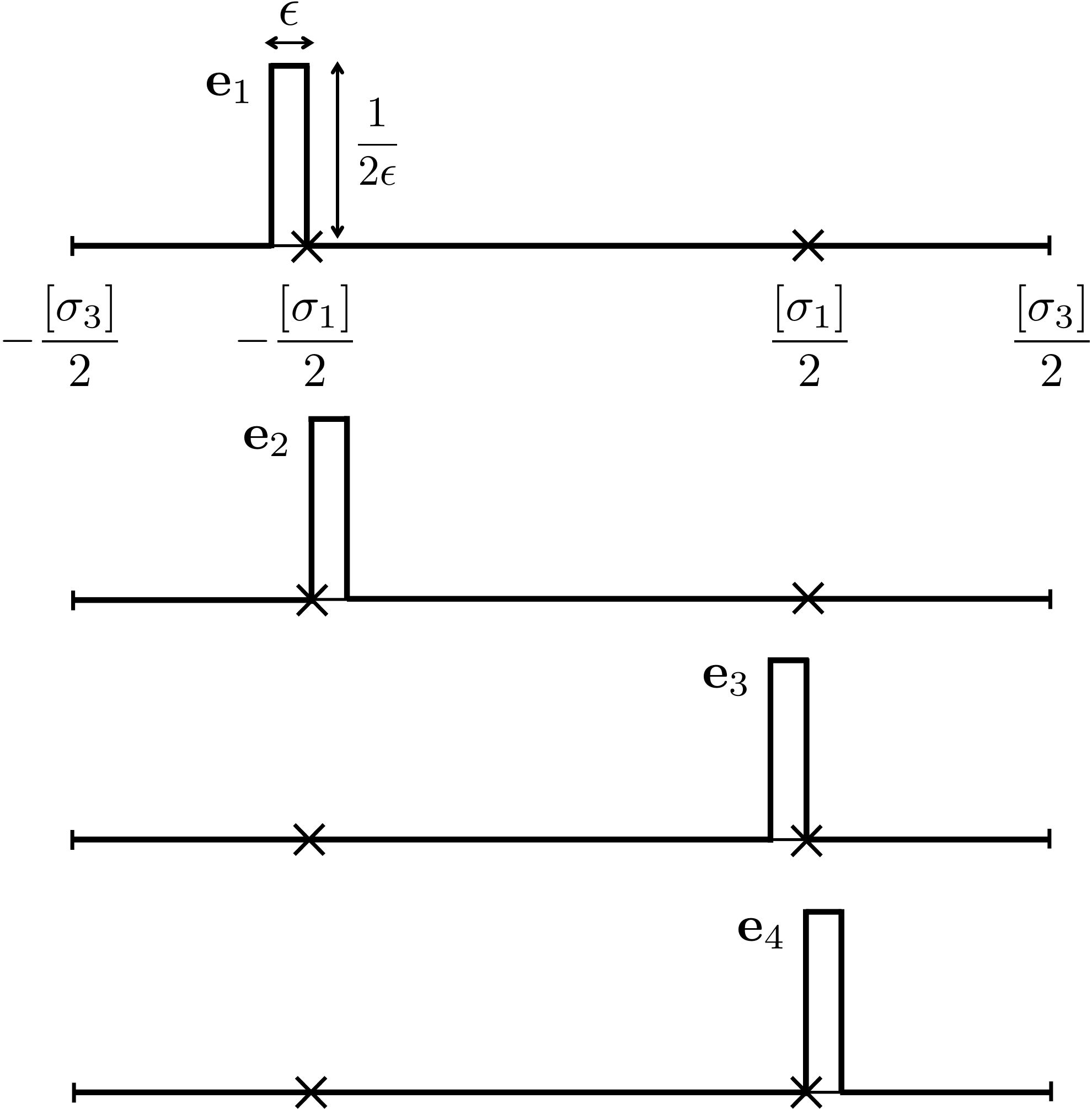}
\caption{
The smeared delta functions localised near the interaction
point (indicated by the crosses) $\bm{e}_i$ with $i=1,\cdots ,4$. 
} 
\label{deltafun}
\end{figure}

One could in general consider any linear combination 
\begin{align}
a_1 \bm{e}_1
+
a_2 \bm{e}_2
+
a_3 \bm{e}_3
+
a_4 \bm{e}_4\ ,
\end{align}
of the four delta function approximations,
$\bm{e}_1, \ldots, \bm{e}_4$, depicted schematically in Fig. \ref{deltafun}.

As explained in section \ref{RSCommutatorComputation} below (\ref{RF1dotw}) 
it is desirable to have
\begin{align}
a_1+a_2+a_3+a_4=0\ ,
\end{align}
in order to eliminate some unwanted contributions in the computation of commutators.

Furthermore it can be shown, using the method of the test functional discussed in 
appendix \ref{RSASmearingTestFunctional}, that
\begin{align}
\bm{e}_1-\bm{e}_2+\bm{e}_3-\bm{e}_4\ ,
\end{align}
gives vanishing contribution as an insertion operator.
Intuitively, this combination vanishes, because it vanishes from 
the perspective of both the $r=1,2$ strings and the $r=3$ string.
In other words, in the limit $\e\rightarrow 0$, the above vanishes as a distribution
both acting on well-behaved periodic functions defined on $I$ and
also on $I_1$ and $I_2$.

Hence we are left with a two-dimensional vector space which 
is spanned by $w$ (\ref{RFDefwFirst})-(\ref{RFDefwSecond}) used in the main text
\begin{align}
w= -\bm{e}_1+ \bm{e}_2 +\bm{e}_3 -\bm{e}_4\ ,
\end{align}
and $v$ defined by
\begin{align}
v= \bm{e}_1+ \bm{e}_2 -\bm{e}_3 -\bm{e}_4\ ,
\label{RFDefv}
\end{align}
or equivalently
\begin{align}
v(\s)=&
\d\left(\s+\frac{[\s_1]}{2}\right)
-\d\left(\s-\frac{[\s_1]}{2}\right)\ ,
\end{align}
\ie
\begin{align}
v^{3m}=&
\frac{2i}{[\s_3]}
\sin{\left(m\pi\frac{[\s_1]}{[\s_3]}\right)}\ .
\end{align}

As explained in detail in appendix \ref{RSASmearingTestFunctional},
$w$ must be used instead of $v$, since this choice assures the vanishing of 
the commutator $[Q_D, P^-]$ to cubic order.

\subsection{Some mathematical properties  of the overlap and the insertions}
\label{RSAPropOverlapInsertion}
In this section we compile mathematical properties of $v$ and $w$ with 
the overlap, which is associated with subtleties related to 
the interaction point.
The formulae in this section are not used in the main text.
We nonetheless present them, since they may play a role in case the need arise to
improve the ansatze presented in the main text. The formulae also 
somewhat clarify the relation of the insertion we used for tensionless strings 
and the insertion used in \cite{RBGS1, RBGSB, RBGS2, RBGS3} for 
tensile superstring field theories.

We will focus on the bosonic sector and denote the 
overlap by $V$ omitting the subscript $B$. 
Analogous properties hold for the fermionic sector as well.

We first introduce another basis-changing matrix (in the opposite direction
compared to (\ref{RFDefA})) defined by
\begin{equation}
x^{3n}= \left( A^{-1}\right)^{3m}{}_{rn} x^{rn}\ ,
\end{equation}
where we hereafter use the convention in which the repeated index $r$ is summed over $1,2$.
We will see below that the notation $A^{-1}$ is somewhat inaccurate.

In~\cite{RBGS1, RBGSB, RBGS2, RBGS3},
the form of the bosonic insertion $Z$ is fixed 
by the requirement that it satisfy
\begin{align}
[Z, x(\s_3)-x(\s_r)]=0\ , \label{RFInsertionConditionGS1}
\\
[Z, p(\s_3)+p(\s_r)]=0\ , \label{RFInsertionConditionGS2}
\end{align}
for $\s\in I_r$ ($r=1,2$) in our notation.
Let us consider a $Z$ which is a linear combination of $x^{rm}$ 
($r=1,2,3$)~\footnote{The arguments below go through with little modification
even if we consider a general linear combination of 
both $x$'s and $p$'s.},
\begin{align}
Z=\sum_{r=1,2,3} \sum_m z_{rm} x^{rm}\ .
\end{align}
We need only consider (\ref{RFInsertionConditionGS2}) which can be re-expressed as
\begin{align}
[Z, p_{rn}+A^{-1\ 3m}{}_{rn} p_{3m}]=0\ ,
\end{align}
or
\begin{align}
[Z, p_{3m}+A^{rn}{}_{3m} p_{rn}]=0\ , \label{RFEqForZ}
\end{align}
depending on the basis we use.

If we employ, say, the latter condition, this implies
\begin{align}
z_{3m}=-A^{rn}{}_{3m} z_{rn}\ .
\end{align}
Hence for any given $z_{rn}$ $(r=1,2)$ we have an insertion 
\begin{align}
Z=z_{rn} \left(A^{rn}{}_{3m}- x^{rn}\right)\ ,
\end{align}
satisfying the condition (\ref{RFEqForZ}).

However, if the $Z$ obtained above acts on the overlap operator $V$, we have
\begin{align}
z_{rn} \left(A^{rn}{}_{3m}- x^{rn}\right)
\prod \d\left(x^{rn}- A^{rn}{}_{3m} x^{3m}\right)=0\ .
\end{align}
Thus all solutions of (\ref{RFEqForZ}) seem to give a vanishing result when 
acting on $V$ and thus cannot be employed as the insertion.

This seemingly paradoxical result could actually have been
anticipated. The conditions, 
(\ref{RFInsertionConditionGS1}) and (\ref{RFInsertionConditionGS2}),
mean that the $r=1,2$ strings and the $r=3$ string are stitched together.
This is the same condition which defines $V$. 
Thus it is natural that the objects 
satisfying (\ref{RFInsertionConditionGS1}) and (\ref{RFInsertionConditionGS2})
annihilate $V$.
The stitching conditions, however, could fail at the interaction point, where we expect 
them to become ill-defined. Thus any object which does not annihilate
$V$ and satisfies (\ref{RFInsertionConditionGS1}) and (\ref{RFInsertionConditionGS2})
is necessarily associated with the interaction point.

This ill-defined nature at the interaction point is reflected in the fact that
the infinite-dimensional matrix $A^{rn}{}_{3m}$ has an eigenvector with zero eigenvalue,
\begin{align}
A^{rn}{}_{3m}v^{3m}=0\ ,
\label{RFAvZero}
\end{align}
where $v$ is defined in (\ref{RFDefv}).
This can be verified directly using the formula
\begin{align}
A^{1m_1}{}_{3 m_3}
=&
(-1)^{m_1}
\frac{1}{
\pi [\s_1] 
\left(
\frac{m_3}{[\s_3]}
-\frac{m_1}{[\s_1]}
\right)
}
\sin\left(
\pi 
\frac{ [\s_1] }{[\s_3]}
m_3
\right)\ ,
\end{align}
\begin{align}
A^{2m_2}{}_{3 m_3}
=&
(-1)^{m_2+1}
\frac{1}{
\pi [\s_2] 
\left(
\frac{m_3}{[\s_3]}
-\frac{m_2}{[\s_2]}
\right)
}
\sin\left(
\pi
\frac{ [\s_1]}{[\s_3]}
 m_3
\right)\ .
\end{align}
A geometrical understanding of this condition is as follows. 
$v$ is a well defined delta function (as a distribution) in the space 
of well-behaved (\ie periodic with no gap) functions on the interval $I$ 
associated with the third string. However it gives vanishing contribution
when acting on well-behaved functions defined on $I_1$, $I_2$ corresponding
to the first and the second strings.

Similarly we have
\begin{align}
A^{-1\ 3m}{}_{rn} w^{rn} =0\ ,
\label{RFAInvwZero}
\end{align}
which again can be verified directly and has a similar geometrical interpretation.

The existence of $v$, $w$ means that the following expression
\begin{equation}
V'=
\prod_m
\d(x^{3m}-(A^{-1})^{3m}{}_{rn}x^{rn})\ ,
\end{equation}
which formally is equivalent to $V$ (up to an overall factor),
is actually subtly different from $V$.

Indeed, it can be shown that whereas
\begin{align}
(x \cdot v) V\ , \ \ \ 
(p \cdot v) V\ , \ \ \
(x \cdot w) V'\ , \ \ \  
(p \cdot w ) V'\ ,   
\label{RFFormallyNonZeroInsertions}
\end{align}
are non-zero, the other combinations are equal to zero
\begin{align}
(x \cdot w) V=0\ , \ \ \ 
(p \cdot w) V=0\ ,  \ \ \ 
(x \cdot v ) V'=0 \ , \ \ \ 
(p \cdot v) V' =0\ .
\label{RFFormallyZeroInsertions}
\end{align}
To understand this, it is instructive to consider the following integral
\begin{equation}
X=\int 
f(\bm{x})
\delta(\bm{x} - A \bm{x'})
g(\bm{x'})
d^3\bm{x}
d^3\bm{x'}\ , 
\label{RFInstructiveIntegral}
\end{equation}
where the matrix A is defined by
\begin{equation}
A=
\begin{bmatrix}
1 & \\
& 1 & \\
& & 0
\end{bmatrix}\ 
\end{equation}
and the ``wave functions'' $f(\bm{x})$ and $g(\bm{x})$ decay sufficiently fast 
for $|\bm{x}| \rightarrow \infty$.
Carrying out the $x', y'$ integral in the usual manner, we obtain
\begin{align}
X&=
\int 
f(x,y,z)
\delta(z-0)
g(x,y,z')
dx dy dz
dz'
\notag
\\
&=
\int 
f(x,y,0)
\left(
\int g(x,y,z') dz'
\right)
dx dy\ . 
\end{align}
We see that in the last expression the integral 
over $z'$ is performed first
and acts only on $g$. Thus the wave function $g$ in the $z'$-direction is averaged over.
Hence whereas inserting $\der_{z'}$ acting on $g$ into (\ref{RFInstructiveIntegral}) gives $0$, the
insertion of $z'$ gives, in general, a non-vanishing contribution. 
On the other hand, the $z$-variable of $f$ is bound firmly to $0$.
Hence in (\ref{RFInstructiveIntegral}) the insertion of $\der_z$ 
acting on $f$ is non-vanishing, while inserting $z$ gives a vanishing
result.

It is interesting to note that when one performs a Fourier transformation
and uses the $p$-representation instead of $x$-representation,
the role of $(V, V'), (A, A^{-1}), (v, w)$ is respectively exchanged
in (\ref{RFFormallyNonZeroInsertions}) and (\ref{RFFormallyZeroInsertions}).
In particular, the momentum representations of $V, V'$ are
\begin{align}
V=&\prod_{m} \d\left(p_{3m}-A^{rn}{}_{3m}p_{rn}\right)\ ,
\\
V'=&\prod_{r=1,2}\prod_n \d\left(p_{rn}-A^{-1\ 3m}{}_{rn}p_{3m}\right)\ ,
\end{align}
up to an overall constant. 

The list of non-zero insertions 
(\ref{RFFormallyNonZeroInsertions}) 
shows that one can choose insertions which satisfy relations such as (\ref{RFEqForZ}) 
but are non-vanishing when acting on the overlap. 
These relations may be useful to construct an ansatz of the cubic vertices
satisfying the superalgebra.
However, there is a caveat associated with the smearing procedure 
explained in appendix \ref{RSASmearingTestFunctional}.

As discussed in appendix \ref{RSASmearingTestFunctional}, it seems that we
need to introduce a smearing of the insertions, say, $\tilde{p}\cdot w=p\cdot \tilde{w}$.
It turns out that 
the identities (\ref{RFAvZero}), (\ref{RFAInvwZero}), and hence (\ref{RFFormallyZeroInsertions}),
become invalid for any finite smearing.
For example,
\begin{align}
\lim_{\e\rightarrow 0} A \tilde{v} \neq 0\ ,
\label{RFAvZeroSmeared}
\end{align}
while it is true that
\begin{align}
A v= 0\ ,
\end{align}
and
\begin{align}
\lim_{\e\rightarrow 0} \tilde{v} =v\ .
\end{align}
Thus the limit involved in the infinite sum over the mode numbers
in the computation of $A v$ does not commute 
with the limit $\e \rightarrow 0$.
This is because  there is a number of order $\sim \frac{1}{\e}$ of terms contributing 
to the sum $A v$, each of which behaves as $\e$. 

Thus, although (\ref{RFFormallyZeroInsertions}) seems to prohibit 
the use of some insertions (since they vanish), the introduction of the smearing 
makes it possible to use them. Also, when smearing is introduced, 
one can ignore the subtle difference between $V$ and $V'$. This is natural since 
the difference is associated with the singularity strictly at the interaction point.

\section{Smearing and test functionals}
\label{RSASmearingTestFunctional}

\subsection{Computation of commutators with smearing}  
\label{RSASmearingCommutator}
In the computations of the commutators described in section \ref{RSCommutatorComputation},
we encounter the multiplication of operators defined at the same point in $\s$-space.
In order to perform the computation in a well defined manner we introduce 
a regularisation of the operators by using a smearing procedure.

Here we will define the smearing procedure and compute, as an example, 
the commutator
\begin{equation}
\left[ Q^{(0)}_{D\dot a A}, Q^{(1)}_{D \dot b B} \right] 
+ \left[ Q^{(1)}_{D\dot a A}, Q^{(0)}_{D \dot b B} \right] 
= \sqrt{2} \epsilon_{\dot a \dot b} C_{AB}  P^{-(1)} 
\label{smearedqdqd}
\end{equation}
using the smeared operators.

We define a smeared version of the momentum density $p(\s)$ by
\begin{equation}
\tilde p(\sigma) 
= \int
 f (\sigma , \sigma') p(\sigma') d\sigma' \ . 
\end{equation}   
One can choose, as the kernel function $f(\s, \s')$, any regularisation 
of the Dirac delta function. 
For definiteness, we choose
\begin{equation}
f(\s, \s') 
= 
\begin{cases}
\frac{1}{2\e} & \text{for}\ \ \sigma - \e \le \sigma' \le \sigma + \e \\
0 & \text{otherwise}
\end{cases} \ ,
\label{kernel1}
\end{equation} 
where
$\e \ll 1$ is the parameter of the smearing.
If $\s$ is close to the interaction point and/or the boundary of the interval on which $\s$ 
is defined,
the above formula should be modified appropriately so that the 
correct periodicity is maintained.

To regularise the terms in the supercharge that are quadratic and cubic in the 
string field one replaces $p(\s)$ in (\ref{RFQDZero}) and (\ref{RFQDOne})  by
its smeared version $\tilde{p}(\s)$, 
\begin{align}
Q^{(0)}_{D\dot aA}
&= \int
\frac{1}{\sqrt2}
\tilde q_{ b A}(\s) \frac{1}{p^+} \e^{bc} \tilde p^\a(\s) \s^\a{}_{c\da}
d\s\ , \label{eq:smearrightaction}
\\
Q^{(1)}_{D\dot a A}
&= f^I{}_{JK} 
\int 
\overline{\phi_{P^+_3}}_I  \notag \\
& \ \ \ \times  
\left( 
(\tilde p_{\alpha} \cdot w) (\sigma^{\alpha b}{}_{\dot a} \tilde d_{bA} \cdot w) 
(p^+)^{\lambda_0}
(P^+_1)^{\lambda_1}
(P^+_2)^{\lambda_2}
(P^+_3)^{\lambda_3} 
\delta (P^+_1 + P^+_2 - P^+_3) V     
\right) \notag \\ 
& \ \ \ \times 
\phi_{P^+_1}{}^J  \phi_{P^+_2}{}^K 
\prod^3_{r=1} 
dP^+_r \mathcal{D}\theta_r  \mathcal{D}x_r 
 \ . 
\label{tildeqd1}
\end{align}  

The computation of $[Q_D^{(0)}, Q_D^{(1)}]$ involves 
\begin{equation}
[ \tilde d_{aA} (\s), \tilde d_{bB} (\s') ] 
= 2p^+ \epsilon_{ab}C_{AB} f' (\s - \s')\ ,
\end{equation} 
where $f'$ is given by the convolution integral, 
\begin{align}
f'(\s,\s') 
&= \int f(\s, \s'') f(\s',\s'')d\s'' \notag \\
&= 
\begin{cases}
- \frac{|\sigma - \sigma'|}{(2\e)^2} + \frac{1}{2\e} & \text{for}\ \ \sigma - 2\e \le \sigma' \le \sigma + 2\e \\
0 & \text{otherwise}
\end{cases},
\label{kernel2}
\end{align} 
satisfying 
\begin{equation}
\int f' (\sigma, \sigma')d\sigma' = 1\ .
\end{equation} 

Using (\ref{kernel2}) as well as (\ref{kernel1}), 
the resulting commutator can be written as
\begin{align}
\left[ Q^{(0)}_{D\dot a A}, Q^{(1)}_{D \dot b B} \right] 
&=2C_{AB} \epsilon_{\dot a \dot b} f^I{}_{JK} 
\int \overline{\phi_{P^+_3}}_I  \notag \\
& \ \ \ \times  
\left( 
(\tilde p_{\alpha}\cdot w) (\tilde p'_{\alpha} \cdot w) 
(p^+)^{\lambda_0}
(P^+_1)^{\lambda_1}
(P^+_2)^{\lambda_2}
(P^+_3)^{\lambda_3} 
\delta (P^+_1 + P^+_2 - P^+_3) V
\right) \notag \\
&\ \ \  \times 
\phi_{P^+_1}{}^J  \phi_{P^+_2}{}^K 
\prod^3_{r=1} 
dP^+_r \mathcal{D}\theta_r  \mathcal{D}x_r\ , 
\label{qd0qd1}
\end{align}
where 
\begin{equation}
\tilde p'_{\alpha} (\sigma) 
= \int f'(\sigma, \s') p_{\alpha} (\sigma')d \sigma' \ ,
\end{equation}
is a smeared version of $p_\a(\s)$.

From (\ref{qd0qd1}), one obtains   
\begin{align}
&P^{-(1)} \notag \\
&= 2\sqrt{2} f^I{}_{JK} 
\int \overline{\phi_{P^+_3}}_I 
\left( 
(\tilde p_{\alpha} \cdot  w)
(\tilde p'_{\alpha} \cdot  w)
(p^+)^{\lambda_0}
(P^+_1)^{\lambda_1}
(P^+_2)^{\lambda_2}
(P^+_3)^{\lambda_3} 
\delta (P^+_1 + P^+_2 - P^+_3) V     
\right) \notag \\ 
& \ \ \ \times 
\phi_{P^+_1}{}^J \phi_{P^+_2}{}^K
\prod^3_{r=1} 
dP^+_r \mathcal{D}\theta_r  \mathcal{D}x_r\ . 
\end{align}

\subsection{Test functionals}
We also occasionally have to deal with complicated expressions involving delta functions 
at the interaction point and delta functionals 
connecting the first and second strings to the third string.
In order to deal with these expressions, 
it is often useful to introduce a set of test functionals
and see how these expressions act on those test functionals.

The test functionals should be sufficiently general.
The set of the test functionals we choose is, for a single string, 
\begin{align}
\phi_k[x]=
e^{-\frac{\a}{4} p^+ \int x(\s)^2 d\s}
\times
e^{i \int k(\s) x(\s) d\s}\ ,
\end{align}
where $k(\s)$, which is a smooth periodic function of $\s$, 
and $\a$ are the parameters of the test functional.

When dealing with string interactions, we use
\begin{align}
\phi_{r}[x_r]=
e^{-\frac{\a}{4} p^+ \int x_r(\s_r)^2 d\s_r}
\times
e^{i \int k_r(\s_r) x_r(\s_r) d\s_r}\ , \qquad r=1, 2, 3 \, ,
\end{align}
where $k_r(\s_r)$ are the parameters of the test functional.
Each $k_r(\s_r)$ is a smooth periodic function defined on 
$\s_r \in \left[-[\s_r]/2, + [\s_r]/2\right]$.

These test functionals are generalised Gaussian
wave packets. This is natural for a tensionless string, which is 
a collection of free particles associated with each value of $\s$. 
The probability distributions 
$|\phi_k|^2$ at each point in $\s$ are uncorrelated.
The distribution corresponds to Gaussian white noise (used for example
in describing Brownian motion). The width of the Gaussian is 
proportional to $\a^{-1}$. 
The factor of $p^+$ in the exponent makes it 
invariant under trivial rescalings of the $\s$ coordinate. 
Also, one has the same Gaussian weight locally for all strings when 
an interaction is considered,
since $p^+$ is common to all three strings due to 
momentum conservation implying $P^+_1+P^+_2=P^+_3$.

We evaluate the expressions by sandwiching them between test functionals.
We first consider basic building blocks in such an analysis.
By standard manipulations of Gaussian integrals
(involving completing the square in the exponent and a shift of the integration contour in the complex plane), we obtain
\begin{align}
\int \overline{\phi_k'}\  p(\s)\ \phi_k \mD x
=&
\int 
\left(\frac{i}{2}\a p^+ x(\s) + k(\s) \right)
e^{-\frac{\a}{2} p^+ \int x^2 d\s+
i \int (k-k') x d\s}
\mD x\notag 
\\
=&
\int 
\left(\frac{i}{2}\a p^+ x + k \right)(\s)
e^{-\frac{\a}{2} p^+ \int 
\left(x-i\frac{k-k'}{\a p^+}\right)^2 d\s}
\mD x
\times
e^{-\frac{\left(k-k'\right)^2}{2\a p^+} d\s}\notag 
\\
=&
\int 
\left(\frac{i}{2}\a p^+ 
\left(x+ i \frac{k-k'}{\a p^+}\right) + k \right)(\s)
e^{-\frac{\a}{2} p^+ \int 
x^2 d\s}
\mD x
\times
e^{-\frac{\left(k-k'\right)^2}{2\a p^+} d\s}\notag 
\\
=&
\frac{k+k'}{2}
(\s)
\times
\mN
e^{-\frac{\left(k-k'\right)^2}{2\a p^+} d\s}\ .
\end{align}
Here $\mN$ is an (infinite) normalisation constant, which may be 
absorbed into the definition of the test functionals $\phi(k)$.

We further have, 
\begin{align}
&\int
\overline{\phi_k'}\ p(\s) p(\s') \phi_k \mD x\notag
\\
=&
\int 
\left(
\left(\frac{i}{2}\a p^+ x + \frac{k+k'}{2} \right)(\s)
\left(\frac{i}{2}\a p^+ x + \frac{k+k'}{2} \right)(\s')
+\frac{\a}{2}p^+\d(\s'-\s)
\right)\notag 
\\
&\times
e^{-\frac{\a}{2} p^+ \int x^2 d\s}
\mD x
\times
e^{-\frac{\left(k-k'\right)^2}{2\a p^+} d\s}\notag 
\\
=&
\left(
\frac{\a}{4} p^+ \d(\s-\s') + \frac{k+k'}{2} (\s)
\frac{k+k'}{2} (\s')
\right)
\times
\mN
e^{-\frac{\left(k-k'\right)^2}{2\a p^+} d\s}\ .
\end{align}

These results can be understood as following from
Wick's theorem with non-zero one point functions.
Namely, we can write
\begin{align}
\langle p(\s) \rangle =& \frac{k+k'}{2} (\s)\ ,
\\
\langle p(\s) p(\s') \rangle =& 
\langle p(\s) \rangle
\langle p(\s') \rangle
+
\contraction{}{p(\s)}{}{p(\s')}
p(\s) p(\s')\notag 
\\
=&
\frac{k+k'}{2} (\s)
\frac{k+k'}{2} (\s')
+
\frac{\a}{4} p^+ \d(\s-\s')\ ,
\end{align}
where we omit the common factor 
$\mN e^{-\frac{\left(k-k'\right)^2}{2\a p^+} d\s}$.

This pattern continues and we have, \textit{e.g.},
\begin{align}
&\langle p(\s) p(\s') p(\s'') \rangle\notag  
\\
=&\langle p(\s) \rangle \langle p(\s') \rangle \langle p(\s'') \rangle
+
\contraction{\langle p(\s) \rangle}{p(\s')}{}{p(\s'')}
\langle p(\s) \rangle p(\s')  p(\s'') 
+
\contraction{}{p(\s)}{\langle p(\s') \rangle}{p(\s'')}
p(\s) \langle p(\s') \rangle p(\s'') 
+
\contraction{}{p(\s)}{}{p(\s')}
p(\s) p(\s') \langle p(\s'') \rangle\notag 
\\
=&
\frac{k+k'}{2} (\s)
\frac{k+k'}{2} (\s')
\frac{k+k'}{2} (\s'')\notag 
\\
&
+
\frac{k+k'}{2} (\s)
\frac{\a}{4} p^+ \d(\s'-\s'')
+
\frac{k+k'}{2} (\s')
\frac{\a}{4} p^+ \d(\s-\s'')
+
\frac{k+k'}{2} (\s'')
\frac{\a}{4} p^+ \d(\s-\s')\ .
\label{RFpppTest}
\end{align}
The use of Wick contractions here is reminiscent of that in the treatment 
of the Brownian motion. 
It may also play a similar role, for the tensionless string theory, to the simplifications via
CFT techniques in ordinary string theory~\cite{RBBPZ, RBFriedanMartinecShenker}.

If we consider a smeared version of $\int p^2 d\s$,
\begin{align}
\int
f(\s,\s') p(\s) p(\s')
d\s d\s'\ , 
\end{align}
for a generic kernel $f(\s, \s')$, 
we have,
\begin{align}
\left\langle
\int
f(\s,\s') p(\s) p(\s')
d\s d\s'
\right\rangle
=
\int
f(\s,\s') 
\frac{k+k'}{2} (\s)
\frac{k+k'}{2} (\s')
d\s
+
\int f(\s,\s) \frac{\a}{4} p^+ d\s\ . 
\end{align}
The first term in this expression has a well defined limit 
when $\e \rightarrow 0$. The second term, on the other hand,  
depends on the choice of the kernel function and generically
is of order $\frac{1}{\e}$. 
It is natural to drop the second term when evaluating these expressions.
This is analogous to 
taking the normal order in tensile string theory.
The steps used in defining a normal ordered form are: ($i$) regularisation of
the product of operators, for instance by point-splitting, ($ii$) evaluation of
matrix elements, ($iii$) subtraction of divergent terms. In our case the 
analog of step ($i$) is smearing, ($ii$) involves the sandwiching by test 
functionals and ($iii$) corresponds to discarding the second term in the 
above formula.

\subsection{Sample computation using test functionals}
\label{RSApSquareOnV}
In order to discuss $[Q_D, P^-]$, 
it is instructive first to 
consider the following expression 
\begin{align}
\int \bphi_{3}
\left(
\int 
{p}_3(\s_3)^2 d\s_3
- \int {p}_1(\s_1)^2 d\s_1
- \int {p}_2(\s_2)^2 d\s_2\right)
V \phi_{1} \phi_{2}
\D x_1 \D x_2 \D x_3\ .
\label{RFpSquaredOnV}
\end{align}
Formal application of (\ref{RFpVZero}) seems to imply that 
this expression vanishes.
However, whether that is true has to be carefully 
examined because of the singularity associated with the multiplication of
$p$'s at the same point in the above formula.

We first introduce the smearing to the above,
\begin{align}
\int \bphi_{3}
\left(
\int 
\tilde{p}_3(\s_3)^2 d\s_3
- \int \tilde{p}_1(\s_1)^2 d\s_1
- \int \tilde{p}_2(\s_2)^2 d\s_2\right)
V \phi_{1} \phi_{2}
\D x_1 \D x_2 \D x_3\ ,
\label{RFptildeSquaredOnV}
\end{align}
where $\tilde{p}_r$ is the smeared momentum density defined for the $r$-th 
string.

For brevity, we introduce $p_{12}(\s)$, defined on the
whole interval $I$, which coincides with $p_r(\s_r)$ for $\s\in I_r$ $(r=1,2)$.
Similarly, we also define $k_{12}(\s)$ out of  $k_1(\s_1)$ and $k_2(\s_2)$.
We have
\begin{align}
&\int 
\tilde{p}_3(\s_3)^2 d\s_3
- \int \tilde{p}_1(\s_1)^2 d\s_1
- \int \tilde{p}_2(\s_2)^2 d\s_2\notag 
\\
=&
\int 
\left(
\tilde{p}_3(\s)^2 
- \tilde{p}_{12}(\s)^2 
\right) d\s\notag 
\\
=&
\int
\left(
p_3(\s)p_3(\s')f_3(\s, \s')
-
p_{12}(\s)p_{12}(\s')f_{12}(\s, \s')
\right)
d\s d\s'\ .
\end{align}
Here $f_3(\s, \s')$ and $f_{12}(\s,\s')$ 
are kernels for the smearing associated with the third 
string and the first-second strings. $f_3$ and $f_{12}$ are different because 
they should obey different periodicity conditions. 
They are the same except when $\s$ and $\s'$ are 
sufficiently close (of the order of the length scale $\e$ of smearing) 
to the interaction point. 

Using (\ref{RFpVZero}) and eliminating the delta functional $V$, (\ref{RFptildeSquaredOnV}) 
becomes
\begin{align}
\int \bphi_{k_3}
\int
p(\s)p(\s')
\left(
f_3(\s, \s')
-
f_{12}(\s, \s')
\right)
d\s d\s'
\phi_{k_{12}}
\D x\ .
\end{align}

Using the short-hand notation introduced in the previous subsection, 
we have
\begin{align}
& \left\langle
\int p(\s)p(\s')(f_3(\s, \s')-f_{12}(\s,\s'))
d\s d\s'
\right\rangle\notag
\\
=&
\left\langle
\int \left(
\frac{\a}{4} p^+\d(\s-\s') + 
k(\s)
k(\s') 
\right)
(f_3(\s, \s')-f_{12}(\s,\s'))
d\s d\s'
\right\rangle\notag
\\
=&
\left\langle
\int \left(
k(\s)
k(\s') 
\right)
\left(
f_3(\s, \s')-f_{12}(\s,\s')
\right)
d\s d\s'
\right\rangle\ ,    
\end{align}
where $k(\s)=\frac{k_{12}(\s)+k_3(\s)}{2}$.
To obtain the last line we used $f_3(\s,\s)=f_{12}(\s,\s)$.

The expression $f_3(\s, \s')-f_{12}(\s,\s')$
is non-zero only if $\s$ is sufficiently near the interaction point.
Examining the behaviour of this expression for each possible case of $\s$ 
(the left/right of the first/second interaction points on $I$) 
and of $\s'$, 
we find that, effectively,
\begin{align}
f_3(\s, \s')-f_{12}(\s,\s')
\sim 
\e \tilde{v}(\s) \tilde{v}(\s')\ ,
\label{RFFOneTwoMinusFThree}
\end{align}
for $\e \ll 1$, where $\tilde{v}(\s)$ is a smeared version of $v(\s)$ 
(a linear combination of delta functions 
having singularities at the vicinity of the interaction point) defined in (\ref{RFDefv}).
Here we omitted an unimportant numerical constant in the RHS.
 
Using this, (\ref{RFptildeSquaredOnV}) becomes finally
\begin{align}
\e v\cdot k v \cdot k\ ,
\end{align}
and thus goes to zero when $\e\rightarrow 0$.
Thus, we have shown that, for the case of (\ref{RFpSquaredOnV}), 
formal manipulations using (\ref{RFpVZero}) are indeed justified
by means of the smearing and the test functionals.

\subsection{$[Q_D, P^-]$ via smearing and test functionals}
\label{RSASmearingTestFunctionalQP}
Now we consider $[P^-, Q_D]=0$. 
There are two contributions in the cubic order, 
$[P^{-(0)}, Q_D^{(1)}]$ and 
$[P^{-(1)}, Q_D^{(0)}]$. The latter can be computed 
in the manner presented in section \ref{RSCommutatorComputation} and 
vanishes.
For the former, one can perform a similar computation
which yields an expression of the following form
\begin{align}
\tilde{p}\cdot w 
\tilde{q}\cdot w 
\int 
\left(
\tilde{p}_3(\s)^2
-
\tilde{p}_{12}(\s)^2
\right) d\s V\ ,
\end{align}
where we omit all unimportant factors.
We have to verify that this expression vanishes 
which needs to be justified using smearing and the test functionals.

Firstly, we notice that the fermionic insertion $\tilde{q}\cdot w$ plays no important role.
It will give a non-singular and non-zero contribution if we introduce appropriate
fermionic contributions in the definition of the test functionals.

Thus we will focus on, by using test functionals,
\begin{align}
&
\int \bphi_3
\tilde{p}\cdot w 
\int 
\left(
\tilde{p}_3(\s)^2
-
\tilde{p}_{12}(\s)^2
\right) d\s 
V
\phi_1 \phi_2
\mD x_1 \mD x_2 \mD x_3\ .
\label{RFpDotwpSquaredOnVTest}
\end{align}

We proceed in a manner similar to the previous subsection.
Using (\ref{RFpVZero}) and eliminating $V$, 
(\ref{RFpDotwpSquaredOnVTest})
can be recast into 
\begin{align}
&
\int \overline{\phi_{k_3}}
\tilde{p}\cdot w 
\int 
p(\s)p(\s')
\left(
f_{3}(\s, \s')
-f_{12}(\s, \s')
\right) 
d\s d\s'
\phi_{k_{12}}
\mD x\ .
\end{align}

In the short-hand notation this becomes, 
using $\tilde{p}\cdot w = p\cdot \tilde{w}$,
\begin{align}
&
\int d\s d\s' d\s''
\tilde{w}(\s'')
\left(
f_3(\s,\s')-f_{12}(\s,\s')
\right)
\left\langle
p(\s'')p(\s)p(\s')
\right\rangle\notag 
\\
= &
\int d\s d\s' d\s''
\tilde{w}(\s'')
\left(
f_3(\s,\s')-f_{12}(\s,\s')
\right)\notag 
\\
\times
&\left(
\langle p(\s) \rangle \langle p(\s') \rangle \langle p(\s'') \rangle
+
\contraction{\langle p(\s) \rangle}{p(\s')}{}{p(\s'')}
\langle p(\s) \rangle p(\s')  p(\s'') 
+
\contraction{}{p(\s)}{\langle p(\s') \rangle}{p(\s'')}
p(\s) \langle p(\s') \rangle p(\s'') 
+
\contraction{}{p(\s)}{}{p(\s')}
p(\s) p(\s') \langle p(\s'') \rangle
\right)\ .
\end{align}
Using (\ref{RFpppTest}), and then $f_{12}(\s,\s)=f_3(\s,\s)$ and (\ref{RFFOneTwoMinusFThree}),
this becomes, omitting an unimportant overall numerical factor,
\begin{align}
\sim\
&
\e 
(\tilde{k}\cdot v)^2 
\tilde{k}\cdot w 
+
\frac{\a}{2}p^+ \e k \cdot \tilde{v} \tilde{v}\cdot \tilde{w}\ .
\end{align}
The first term vanishes in the limit $\e\rightarrow 0$.
This is also the case for the second term because $v\cdot w=0$.

An important point here is that had we chosen to construct the ansatz in terms
of $v$, the second term would have become $2 \e \tilde{k}\cdot v \tilde{v}\cdot \tilde{v}$.
This gives a finite contribution, since $\tilde{v}\cdot \tilde{v} \sim \frac{1}{\e}$.
This would be inconsistent with the superalgebra. 
This justifies our use of $w$, rather than $v$, for 
insertions in our ansatz of the dynamical supercharge.

We also notice that formal application of (\ref{RFpVZero}) 
to~(\ref{RFpDotwpSquaredOnVTest})
yields zero automatically irrespective of 
the choice of $v$ or $w$ in the insertion.
The smearing and the test functional method we developed 
show that such formal application is not allowed due to the singularity
associated with multiplication of $p(\s)$'s at the same point.
A contribution to the commutator $[P^-, M^{-\a}]$ in light-cone gauge bosonic 
string theory arising by essentially the same mechanism is discussed 
in~\cite{RBMandelstamLorentz}.
There the critical dimension $d=26$ follows from requiring 
that the contribution vanishes.

\end{document}